\documentclass[11pt]{article}
\pdfoutput=1

\usepackage{jheppub}
\usepackage{bigints}

\usepackage{tabularx}

\usepackage{tikz}
\usepackage[bb=boondox]{mathalfa}

\usepackage{ae}
\usepackage[T1]{fontenc}
\usepackage{url}

\usepackage{tabularx}

\usepackage{graphicx}
\usepackage{dsfont}
\usepackage{setspace}
\usepackage{amsfonts,amsmath,amsthm,amssymb,amsbsy}
\usepackage{longtable}
\usepackage[latin1]{inputenc}
\usepackage{array}
\usepackage{color}
\usepackage{needspace}
\usepackage[numbers]{natbib}
\usepackage{colortbl}
\usepackage{arydshln}
\usepackage{appendix}

\usepackage{enumitem}

\usepackage{fancyhdr}

\newcommand{\msf}[1]{\mathsf{#1}}
\newcommand{\idM}{\mathds{1}}
\newcommand{\<}{\langle}
\renewcommand{\>}{\rangle}

\newcommand{\dt}[1]{\dot{#1}}
\newcommand{\tl}[1]{\tilde{#1}}
 	
\newcommand{\diff}{\mathop{}\!\mathrm{d}}	
	
\newcommand{\Le}{\textup{\protect\scalebox{-1}[1]{L}}}

\title{Amplituhedra, and Beyond}

\author[1,2]{Livia Ferro,}\emailAdd{livia.ferro@lmu.de}
\author[2]{Tomasz \L ukowski,}\emailAdd{t.lukowski@herts.ac.uk}

\affiliation[1]{Arnold--Sommerfeld--Center for Theoretical Physics,\\ Ludwig--Maximilians--Universit\"at, \\ Theresienstra\ss e 37, 80333 M\"unchen, Germany }
\affiliation[2]{School of Physics, Astronomy and Mathematics, \\ University of Hertfordshire, \\  Hatfield, Hertfordshire, AL10 9AB, United Kingdom}

\abstract{
This review is a primer on recently established geometric methods for observables in quantum field theories. The main emphasis is on amplituhedra, i.e.~geometries encoding scattering amplitudes for a variety of theories. These pertain to a broader family of geometries called positive geometries, whose basics we review. We also describe other members of this family that are associated with different physical quantities and briefly consider the most recent developments related to positive geometries. Finally, we discuss the main open problems in the field. This is a Topical Review  invited by Journal of Physics A: Mathematical and Theoretical.}

\begin{document}

\begin{flushright}
{\small LMU-ASC 31/20}
\end{flushright}

\maketitle

%%%%%%%%%%%%%%%%%%%%%%%%%%%%%%%%%%%%%%%%%%%%%%%%%%%%%
%%%%%%%%%%%%%%%%%%%%%%%%%%%%%%%%%%%%%%%%%%%%%%%%%%%%%
%%%%%%%%%%INTRODUCTION%%%%%%%%%%%%%%%%%%%%%%%%%%%%%%%
%%%%%%%%%%%%%%%%%%%%%%%%%%%%%%%%%%%%%%%%%%%%%%%%%%%%%
%%%%%%%%%%%%%%%%%%%%%%%%%%%%%%%%%%%%%%%%%%%%%%%%%%%%%

\section{Introduction}

Our understanding of quantum field theories is rapidly changing. 
In recent years we have witnessed the birth of a new paradigm for studying certain 
physical quantities. This development has been mainly driven by
the investigation of scattering amplitudes, with the discovery of new fascinating geometric constructions underlying them. 
In this geometric description, the scattering amplitudes -- and, more generally, the physical quantities -- are
encoded in particular bounded regions, with appropriate properties on their boundaries. %
Despite differing in detail, these constructions rely on a common mathematical structure called \emph{positive
geometry}. Nowadays, positive geometries are appearing for a wide spectrum of theories and quantities. These
range from scattering amplitudes to correlation functions and cosmological observables.
A positive geometry is defined as a real, oriented,
closed geometry with boundaries of all co-dimension. Each boundary is again a positive geometry. The
most important feature is that every positive geometry has a unique differential form, called the canonical form, with logarithmic singularities along all boundaries. Moreover, the residue along a
boundary is given by the canonical form on the boundary.
For physically relevant positive geometries, the canonical form is a physical quantity. Locality and unitarity manifest themselves by the fact that, when we approach one of the boundaries, the quantity which we study appropriately factorises into smaller pieces.
This is a recurring pattern in high-energy physics, where it is common to use recursion
relations to construct more complicated objects from simpler ones.

What we now call ``positive geometries" have made their first appearance in supersymmetric gauge
theories: the prime example, the amplituhedron \cite{Arkani-Hamed:2013jha}, computes tree- and loop-level (integrands) of $n$-point
amplitudes of any helicity sector in maximally supersymmetric Yang-Mills theory  in the
planar limit in momentum twistor space. 
Few years later, for the same theory, the momentum amplituhedron was defined \cite{Damgaard:2019ztj}, which computes the tree-level amplitudes directly in spinor helicity space.
Nowadays, we have found examples of
such structures for scattering amplitudes and other observables in a variety of theories. For
instance, the kinematic associahedron \cite{Arkani-Hamed:2017mur} computes tree-level amplitudes for the bi-adjoint $\phi^3$ theory. 
On-going works have extended kinematic and worldsheet
associahedra to loop-level amplitudes in $\phi^3$ theory, generalised worldsheet and string
integrals, and uncovered deep connections with mathematics such as cluster algebras,
tropical geometry and convex geometry. %
These geometrical constructions have appeared very recently also in cosmology \cite{Arkani-Hamed:2017fdk}: the
cosmological polytope gives a connection between positive geometries and the wave
function of the universe, analogously to the one seen for scattering amplitudes. Positive geometries are arising also in more general conformal field theories (CFT), beyond
maximally supersymmetric Yang-Mills theory. A novel geometric interpretation of the conformal bootstrap equation has been
discovered, which leads to new insights into the four-point functions in CFT \cite{Arkani-Hamed:2018ign}.

In this review, we present a self-contained description of ``Amplituhedra" and, more generally, of positive geometries which underlie physical quantities.
In section \ref{sec:positive.geometry} we start by introducing the mathematical notion of positive geometries, explaining how to determine the canonical form and giving few simple examples. In particular, we define the positive Grassmannian, which provides an auxiliary space used to define amplituhedra later on. Section \ref{sec:amplituhedra} focuses on the positive geometries for  (planar) $\mathcal{N}=4$ super Yang-Mills (sYM). 
In particular we will describe the amplituhedron, the correlahedron and the momentum amplituhedron, 
which is relevant for tree-level amplitudes directly in spinor helicity space. We follow in section \ref{sec:associahedron} with amplituhedra for bi-adjoint cubic scalar theory and in section \ref{sec:other} with positive geometries relevant for observables other than scattering amplitudes. Finally, in section \ref{sec:recent} we summarise recent advances related to positive geometries, including relations to string theory and tropical geometry. We devote the appendix to introduce some relevant notation.

%%%%%%%%%%%%%%%%%%%%%%%%%%%%%%%%%%%%%%%%%%%%%%%%%%%%%%%%
%%%%%%%%%%%%%%%%%%%%%%%%%%%%%%%%%%%%%%%%%%%%%%%%%%%%%%%%%%%%
%%%%%%%%%%%%%%%%POSITIVE GEOMETRIES%%%%%%%%%%%%%%%%%%%%%%%%%
%%%%%%%%%%%%%%%%%%%%%%%%%%%%%%%%%%%%%%%%%%%%%%%%%%%%%%%%%%%	
%%%%%%%%%%%%%%%%%%%%%%%%%%%%%%%%%%%%%%%%%%%%%%%%%%%%%%%%%%%%%

\section{Positive Geometries}
\label{sec:positive.geometry}
We start our survey by providing the definition of positive geometries, somehow reversing the chronological order of the developments described in this review. This has however the advantage of allowing us to discuss all objects in the following sections in a unified framework. We provide here a simplified description of this class of geometries and we refer the reader to \cite{Arkani-Hamed:2017tmz} for a precise definition. Importantly, there are two main ingredients that we need to specify in order to define a positive geometry: a geometric space and a rational differential form. The space is given by a pair: a complex variety $X$ which provides an ambient space, inside which we define a subset $X_{\geq0}$ of its real slice. Then the differential form $\Omega(X,X_{\geq0})$ needs to be meromorphic on $X$ and to behave logarithmically when approaching any boundary of $X_{\geq0}$. Moreover, when we restrict the differential form $\Omega(X,X_{\geq0})$ to any boundary of $X_{\geq0}$ by performing an appropriate residue operation, we obtain the canonical differential form for this boundary.

%%%%%%%%%%%%%%%%%%%%%%%%%%%%%%%%%%%%%%%%%%%%%%%%%%%%%%%%%%
%%%%%%%%%%%%%%%%%%%%%%%%%%%%%%%%%%%%%%%%%%%%%%%%%%%%%%%%%%
%%%%%%%%%%%%%%%%%%%%%%%%%%%%%%%%%%%%%%%%%%%%%%%%%%%%%%%%%%

\subsection{Definition}
Positive geometries naturally live in complex projective spaces, which we denote as $\mathbb{P}^N$, and their real parts $\mathbb{P}^N(\mathbb{R})$. We define $X$ to be a complex projective algebraic variety of complex dimension $D$ and we denote by $X_{\geq 0} \subset X(\mathbb{R})$ an oriented set of real dimension $D$.
 A $D$-dimensional {\it positive geometry} is a pair $(X, X_{\geq 0})$ equipped with a unique non-zero differential $D$-form $\Omega(X, X_{\geq 0})$ satisfying the following recursive axioms:
\begin{itemize}
\item For $D = 0$ we have  that $X=X_{\geq 0}$ is a single real point and $\Omega(X, X_{\geq 0})=\pm 1$ depending on the orientation of $X_{\geq 0}$.
\item  For $D > 0$ we have that every boundary component $(C, C_{\geq 0})$ of $(X, X_{\geq 0})$ is a positive geometry of dimension $D-1$. Moreover, the form $\Omega(X, X_{\geq 0})$ is constrained by the residue relation 
\begin{equation}
\mbox{Res}_C\, \Omega(X, X_{\geq 0}) = \Omega(C, C_{\geq 0})\,,
\end{equation}
 along every boundary component $C$, and has no singularities elsewhere.
\end{itemize}
The {\it residue} operation $\mbox{Res}_C$ for a meromorphic form $\omega$ on $X$ is defined in the following way: suppose $C$ is a subvariety of $X$ and $z$ is a holomorphic coordinate whose zero set $z = 0$ parametrises $C$. Denote as $u$ the remaining holomorphic coordinates. Then a simple pole of $\omega$ at $C$ is a singularity of the form
\begin{equation}
\omega(u, z) = \omega'(u) \wedge \frac{dz}{z}
+ \ldots\,,
\end{equation}
where the ellipsis denotes terms smooth in the small $z$ limit, and $\omega'(u)$ is a non-zero meromorphic form on the boundary component. One defines
\begin{equation}
\text{Res}_C\, \omega := \omega'\,.
\end{equation}
If there is no such simple pole then one defines the residue to be zero.

%%%%%%%%%%%%%%%%%%%%%%%%%%%%%%%%%%%%%%%%%%%%%%%%%%%%%%%%%%
%%%%%%%%%%%%%%%%%%%%%%%%%%%%%%%%%%%%%%%%%%%%%%%%%%%%%%%%%%
%%%%%%%%%%%%%%%%%%%%%%%%%%%%%%%%%%%%%%%%%%%%%%%%%%%%%%%%%% 
 
\subsection{Positive Geometries in Physics}
\label{sec:physics}
When exploring positive geometries from the point of view of physics, we are interested in defining a region inside the kinematic space relevant for the problem at hand. Often this region can be determined by studying the physical properties of the observables and, in particular, by studying the structure of their singularities. For example, in the case of scattering amplitudes it is known that they diverge when particular combinations of momenta vanish. This determines the boundary structure of the sought-after region and gives strong indications  to determine 
the complete geometry and, afterwards, its canonical form.

Positive geometries provide a broad class of, yet unexplored, geometries. In the physics context we will however restrict our attention to a narrower class of objects and we will distinguish two types of positive geometries relevant for applications in high-energy physics: 
\begin{itemize}
\item In the first class of geometries we will have $X=\mathbb{P}^D$ and $X_{\geq 0}$ will be defined as a collection of linear inequalities and therefore will have properties of a convex polyhedron. Examples include projective simplices, projective embeddings of associahedra relevant for the $\phi^3$ theory, cosmological polytopes, cyclic polytopes and positive geometries for conformal field theories.
\item The second class of objects is related to Grassmannian spaces\footnote{Since a projective space is also an example of a Grassmannian space, then some members of this family of geometries, \emph{e.g.} cyclic polytopes, will also belong to the first class.} and can be pictured as a curvy version of convex polyhedra. It includes positive Grassmannians, the amplituhedron, the momentum amplituhedron and the correlahedron. The only member in this class which is proven to be a positive geometry is the positive Grassmannian; however, there is substantial evidence that also amplituhedra satisfy the axioms of positive geometries. In particular, using physics motivations, explicit expressions for the canonical forms $\Omega(X,X_{\geq0})$ of the amplituhedron and  momentum amplituhedron can be found using the Britto-Cachazo-Feng-Witten (BCFW) recursion relations  \cite{Britto:2004ap,Britto:2005fq}.
\end{itemize}

%%%%%%%%%%%%%%%%%%%%%%%%%%%%%%%%%%%%%%%%%%%%%%%%%%%%%
%%%%%%%%%%%%%%%%%%%%%%%%%%%%%%%%%%%%%%%%%%%%%%%%%%%%%
%%%%%%%%%%%%%%%%%%%%%%%%%%%%%%%%%%%%%%%%%%%%%%%%%%%%%

\subsection{Canonical Forms and How To Find Them}
\label{sec:findforms}
In order to check whether a given pair $(X,X_{\geq0})$ is a positive geometry we need to have an efficient way to find rational differential forms $\Omega(X,X_{\geq0})$ associated to them. There are various different ways to determine such forms and we list below some of the most commonly used: 
\begin{itemize}
\item
{\bf Triangulations:} in this approach the geometry is divided into smaller pieces for which the canonical forms are known. There are two types of triangulations: triangulations introducing spurious boundaries, or the so-called local triangulations, for which we have only physical singularities but need to introduce additional points. 
Each element of a triangulation of the first type has non-physical singularities on spurious boundaries however, since the canonical forms for each smaller geometry are logarithmic, then one gets cancellations on each spurious boundary, leading to a differential form with singularities only on the true boundaries of the positive geometry.

Finding triangulations of a given positive geometry is an interesting, and sometimes difficult, task on its own. For projective convex polytopes there is a range of known algorithms  to accomplish it. On the other hand, for positive geometries in Grassmannian spaces it is often possible to exploit the structure of the positive Grassmannian and arguments from physics to find their triangulations.

\item
{\bf Push-forwards:} it is often possible to find a simpler positive geometry which can be mapped bijectively to (subsets of) more complicated positive geometries. Then we can use such a map to push-forward the known canonical form of the simpler positive geometry to obtain the canonical form for the more complicated one.
 
\item 
{\bf Integral representations -- dual geometry:} for projective convex polytopes it is possible to find their dual polytopes using projective duality. The canonical differential form can then be obtained from the volume of the dual geometry. This justifies the use of the notion of {\it volume form} to indicate a canonical form. For positive geometries in Grassmannian spaces, the notion of a ``dual" is yet to be understood, but some work in this direction was done in \cite{Ferro:2015grk,Arkani-Hamed:2017tmz}. 

\item
{\bf Direct construction from poles and zeros:} knowing that the singularities of the canonical form are located solely at the boundaries of the space $X_{\geq0}$, it allows us to write $\Omega(X,X_{\geq0})$ as a rational function with known denominator factors and a polynomial function in the numerator. In various cases this numerator can be completely fixed by imposing the residue constraints from the definition of the positive geometries \cite{Arkani-Hamed:2014dca}.

\item
{\bf Integral representations -- contour integrals:} we will recall in the following that it is possible to write the canonical forms for the amplituhedron and the momentum amplituhedron  as contour integrals over a Grassmannian space, and for the cosmological polytope as a contour integral over the projective space. In simple cases, the positivity completely fixes the integration contour and allows one to write the volume form as a sum of appropriate residues of this integral.

\end{itemize}

%%%%%%%%%%%%%%%%%%%%%%%%%%%%%%%%%%%%%%%%%%%%%%%%%%%%%%%%%%
%%%%%%%%%%%%%%%%%%%%%%%%%%%%%%%%%%%%%%%%%%%%%%%%%%%%%%%%%
%%%%%%%%%%%%%%%%%%%%%%%%%%%%%%%%%%%%%%%%%%%%%%%%%%%%%%%%%%

\subsection{Basic Examples of Positive Geometries}
We start our exploration of positive geometries by giving few basic examples. In particular, we introduce the general notion of projective polytopes, which include positive geometries  belonging to the first class we mentioned earlier. We also recall the definition of the positive Grassmannian and its properties, which will be relevant in our later explorations of amplituhedra.

%%%%%%%%%%%%%%%%%%%%%%%%%%%%%%%%%%%%%%%%%%%%%%%%%%%%%%%%

\subsubsection{Projective Polytopes}
Positive geometries provide a class of spaces which are generically quite complicated;  however, they also include simple and familiar objects. The simplest examples of positive geometries are simplices, or rather their embedding into the projective space. One defines a {\it projective $m$-simplex} $(\mathbb{P}^m, \Delta)$ as a positive geometry in $\mathbb{P}^m$ cut out by exactly $m+1$ linear inequalities. If we take $Y \in \mathbb{P}^m$ to be a point in projective space with homogeneous components $Y^A$ indexed by $A = 0, 1,\ldots, m$, then any linear inequality in projective space is of the form $Y\cdot W:=Y^A W_A \geq 0$, where $W \in \mathbb{R}^{m+1}$ is a dual vector with components $W_A$. The projective simplex is therefore the set
\begin{equation}
\Delta = \{Y \in \mathbb{P}^m(\mathbb{R}) \,|\, Y \cdot W_i \geq 0 \text{, for } i = 1,\ldots , m+1\}\,.
\end{equation}
Here the $W_i$'s are projective dual vectors corresponding to the facets of the simplex.
Every boundary of a projective simplex is again a projective simplex, it is therefore easy to see that projective simplices satisfy the axioms of a positive geometry. Moreover, we can write down an explicit form of the canonical differential form $\Omega(\mathbb{P}^m,\Delta)$ in terms of the vertices or, equivalently, in terms of the facets of $\Delta$. Let $Z_i\in \mathbb{R}^{m+1}\setminus \{0\}$ denote the vertices of $\Delta$ for $i=1,\ldots,m+1$. Then the canonical form is
\begin{equation}
\Omega(\mathbb{P}^m,\Delta)=\frac{\langle Z_1 Z_2\ldots Z_{m+1}\rangle^m\langle Yd^mY\rangle}{m!\,\langle Y Z_1\ldots Z_m\rangle\,\langle Y Z_2\ldots Z_{m+1}\rangle\ldots \langle YZ_{m+1}\ldots Z_{m-1}\rangle} \,,
\end{equation}
where we denoted
\begin{equation}
\frac{1}{m!}\langle Yd^m Y\rangle=\sum_{A=1}^{m+1}(-1)^{A}Y^{A}dY^1\wedge\ldots\wedge \widehat{dY^A}\wedge \ldots \wedge dY^{m+1}\,,
\end{equation}
and introduced the brackets $\langle\rangle$ which are maximal minors of the matrix $(Y,Z_1,Z_2,\ldots,Z_n)$.

More generally, we can define {\it convex projective polytopes} with vertices $Z_1, Z_2, \ldots , Z_n \in \mathbb{R}^{m+1}\setminus\{0\}$. We denote by $Z$ the $n \times (m + 1)$ matrix whose rows are given by the $Z_i$ and assume that $Z$ is a positive matrix, \emph{i.e.}~a matrix with all maximal minors positive. We define $\mathcal{A} := \mathcal{A}(Z) := \mathcal{A}(Z_1, Z_2, \ldots , Z_n) \subset \mathbb{P}^m(\mathbb{R})$ to be the convex hull of points $Z_1,\ldots,Z_n$
\begin{equation}
\mathcal{A} = \text{Conv}(Z) = \text{Conv}(Z_1, \ldots, Z_n) := \left\{\sum_{i=1}^n
c_i Z_i \in \mathbb{P}^m(\mathbb{R})\, |\, c_i \geq 0, i = 1, \ldots , n\right\} \,.
\end{equation}
We usually restrict to the case where the points $Z_1, \ldots , Z_n$ are all vertices of $\mathcal{A}$.
The polytope $\mathcal{A}$ is called a convex projective polytope
and it is easy to check that it defines a positive geometry. This follows from the fact
that every polytope $\mathcal{A}$ can be triangulated by projective simplices. The canonical form $\Omega(\mathbb{P}^m,\mathcal{A})$ of
a projective polytope can then be found as the sum of canonical forms for the projective simplices triangulating it.

Finally, we observe that every convex polytope in $\mathbb{R}^m$ can be uplifted to a projective polytope in the following way: a convex polytope $A$ can be described as the convex span of some number of vertices $z_1,\ldots,z_n$, where $z_i\in \mathbb{R}^m$. Then we can embed it into a projective space $\mathbb{P}^m$ by constructing the points
\begin{equation}
Z_i=\left(\begin{tabular}{c}
1\\$z_i$
\end{tabular}\right),
\end{equation} 
up to a rescaling. The projective polytope associated to $A$ is then $\mathcal{A}(Z_1,\ldots,Z_n)$.

%%%%%%%%%%%%%%%%%%%%%%%%%%%%%%%%%%%%%%%%%%%%%%%%%%%%%%

\subsubsection{Positive Grassmannian}
A more involved example of  positive geometry is given by the positive Grassmannian -- a generalisation of a projective simplex. This positive geometry plays also a crucial role in the definition of amplituhedra later on, which in turn can be viewed as generalisations of projective polytopes.

We start by defining the {\itshape (real) Grassmannian} $G(k,n)$ (for $0\le k \le n$) which is the space of all $k$-dimensional subspaces of $\mathbb{R}^n$.  An element of $G(k,n)$ can be viewed as a $k\times n$ matrix of rank $k$ modulo invertible row operations, whose rows give a basis for the $k$-dimensional subspace.
We define $[n]=\{1,\dots,n\}$, and denote by $\binom{[n]}{k}$ the set of all $k$-element subsets of $[n]$. Given a Grassmannian element $V\in G(k,n)$ represented by a $k\times n$ matrix $A$, for $I\in \binom{[n]}{k}$, we denote by $p_I(V)$ the $k\times k$ minor of $A$ constructed using the columns in $I$. The $p_I(V)$ do not depend on our choice of the matrix $A$ (up to simultaneous rescaling by a nonzero constant), and are called the {\itshape Pl\"{u}cker coordinates} of $V$.

We say that $V\in G(k,n)$ is {\itshape totally nonnegative} if all Pl\"{u}cker coordinates $p_I(V)\ge 0$ are nonnegative for all $I\in\binom{[n]}{k}$.  
	The set of all totally nonnegative $V\in Gr(k,n)$ is the {\it totally nonnegative Grassmannian} $G_{+}(k,n)$, which we will often refer to as the \emph{positive Grassmannian}.
	For $M\subseteq \binom{[n]}{k}$, we take $S_{M}$ to be
the set of $V\in G_+(k,n)$ with the prescribed collection of Pl\"{u}cker coordinates strictly positive, \emph{i.e.}~$p_I(V)>0$ for all $I\in M$, and the remaining Pl\"{u}cker coordinates equal to zero. We call $S_M$ a \emph{positroid cell} of $G_+(k,n)$. As shown in \cite{Postnikov:2006kva}, the positroid cells of $G_+(k,n)$ are in bijection  with various combinatorial objects, including  \emph{decorated permutations} $\pi$ on $[n]$ with $k$ anti-excedances,  \emph{\Le -diagrams} $D$ of type $(k,n)$, and equivalence classes of \emph{reduced plabic graphs} $G$ of type $(k,n)$.
The positive Grassmannian $G_+(k,n)$ is a $k\times (n-k)$ dimensional space, with an interesting and well-understood boundary structure including positroid cells of all dimensions, which is known to be homeomorphic to a ball \cite{Galashin:2017onl}.   
 
The positive Grassmannian has started to play a prominent role in the development for scattering amplitudes after it was realised that the plabic graphs classified by Postnikov \cite{Postnikov:2006kva} correspond to on-shell diagrams obtained by solving the BCFW recursion relations in planar $\mathcal{N}=4$ sYM theory. The latter allow one to find the amplitude as a sum of (on-shell) graphs with trivalent vertices of two types, corresponding to two three-particle scattering amplitudes $A_{3,1}$ and $A_{3,2}$. Using this relation, the tree-level amplitude  $A^{\text{\tiny tree}}_{n,k}$ corresponds to a particular collection of positroid cells in the positive Grassmannian $G_+(k,n)$. A comprehensive study of the relation between positive Grassmannians and scattering amplitudes can be found in \cite{ArkaniHamed:2012nw}.

%%%%%%%%%%%%%%%%%%%%%%%%%%%%%%%%%%%%%%%%%%%%%%%
%%%%%%%%%%%%%%%%%%%%%%%%%%%%%%%%%%%%%%%%%%%%%%%
%%%%%%%%%%%%%%AMPLITUHEDRA%%%%%%%%%%%%%%%%%%%%
%%%%%%%%%%%%%%%%%%%%%%%%%%%%%%%%%%%%%%%%%%%%%%%
%%%%%%%%%%%%%%%%%%%%%%%%%%%%%%%%%%%%%%%%%%%%%%%

\section{Amplituhedra for \texorpdfstring{$\mathcal{N}=4$}{} sYM Theory}
\label{sec:amplituhedra}
After having presented simple examples of positive geometries and their properties, we are now ready to study the first example of such geometries relevant to physics. The focus of this section is on $\mathcal{N}=4$ sYM and we describe three geometries relevant for scattering amplitudes in this theory: the amplituhedron $\mathcal{A}_{n,k}^{(4)}$ and the loop amplituhedron $\mathcal{A}_{n,k}^{\ell\text{\tiny -loop}}$, which are defined on the momentum twistor space, and the momentum amplituhedron $\mathcal{M}_{n,k}^{(4)}$ which is defined on the spinor helicity space. Moreover, we recall the definition of the correlahedron which is the geometry encoding the stress-energy correlators in planar $\mathcal{N}=4$ sYM. 
The definition of amplituhedra has been also extended beyond the cases relevant to physics: a general definition for the amplituhedron $\mathcal{A}_{n,k}^{(m)}$ was introduced in \cite{Arkani-Hamed:2013jha}, while for the momentum amplituhedron $\mathcal{M}_{n,k}^{(m)}$, for even $m$, in \cite{Lukowski:2020dpn}. These are positive geometries which often serve as a playground for testing the ideas for the physical case $m=4$. In particular, much is known for $m=1,2$ as we summarise in the following sections.

For each positive geometry we will follow a common template in describing its properties. We start by providing a definition, or in many cases few equivalent definitions which highlight different properties of the same geometry. Using these definitions we explain the structure of their boundaries which is necessary to determine whether they are positive geometries. Next, we describe known methods for finding the canonical forms and, if available, provide explicit expressions for them. In many cases no such explicit expressions are known and one needs to refer to a case-by-case study. Finally, we explain how a relevant physical observable is encoded by each positive geometry.

Before we delve into the world of amplituhedra, we remind the reader of a few basic facts about scattering amplitudes for $\mathcal{N}=4$ sYM, which will set the stage and allow us to compare the results which we obtain from positive geometries with known results for amplitudes obtained using standard methods. We also comment on the symmetries of scattering amplitudes.

%%%%%%%%%%%%%%%%%%%%%%%%%%%%%%%%%%%%%%%%%%%%%%%%%%%%%%%%%
%%%%%%%%%%%%%%%%%%%%%%%%%%%%%%%%%%%%%%%%%%%%%%%%%%%%%%%%
%%%%%%%%%%%%%%%%%%%%%%%%%%%%%%%%%%%%%%%%%%%%%%%%%%%%%%%%

\subsection{Scattering amplitudes in \texorpdfstring{$\mathcal{N}=4$}{} sYM}
Scattering superamplitudes in $\mathcal{N}=4$ sYM are defined for on-shell chiral superfields, which collect the on-shell multiplet into a single object by means of the Grassmann-odd variables $\eta_A$ with $A=1,\dots,4$:
\begin{equation}
\label{superfield}
\Phi=G^+ +\eta_A \, \Gamma^A +\frac{1}{2!}\eta_A\eta_B \, S^{AB} +\frac{1}{3!}\eta_A\eta_B\eta_C \, \epsilon^{ABCD} \bar{\Gamma}_{D}+\frac{1}{4!}\eta_A\eta_B\eta_C\eta_D \epsilon^{ABCD}\, G^-,
\end{equation}
with positive (resp. negative) helicity gluons $G^+$ (resp. $G^-$), fermions $\Gamma,\bar\Gamma$ and scalars $S$. 
A generic $n$-particle superamplitude $\mathcal{A}_n = \mathcal{A}_n(\Phi_1,\Phi_2,\ldots,\Phi_n)$ can be expanded in terms of helicity sectors
\begin{align}
\mathcal{A}_n&=A_{n,2} + A_{n,3} + \cdots + A_{n,n-2}, & &(n\ge 4)\,,
\end{align}
where $A_{n,k}$ is the superamplitude for the N$^{k-2}$MHV sector and has Grassmann degree $4k$, \emph{i.e.}~it is proportional to $\eta^{4k}$. Each amplitude $A_{n,k}$ can be further expanded in the coupling constant
\begin{equation}\label{eq:perturbation}
A_{n,k}=A_{n,k}^{\text{\tiny tree}}+\sum_{\ell>0}\lambda^\ell A_{n,k}^{\ell\text{\tiny -loop}}\,,
\end{equation}
where $\lambda$ is the t'Hooft coupling. The positive geometries which encode the tree amplitudes $A_{n,k}^{\text{\tiny tree}}$ are $\mathcal{A}_{n,k-2}^{(4)}$ and $\mathcal{M}_{n,k}^{(4)}$, while the loop amplituhedron $\mathcal{A}_{n,k}^{\ell\text{\tiny -loop}}$ encodes the integrands for  $A_{n,k-2}^{\ell\text{\tiny -loop}}$.

Importantly, the amplitudes are functions of kinematic variables and over the years various kinematic spaces have been used to encode them. The most popular ones are momenta and polarisation vectors, spinor helicity variables, twistor or momentum twistor variables, and their appropriate supersymmetric extensions, see Appendix \ref{app:spaces} for a detailed descriptions of these variables.

%%%%%%%%%%%%%%%%%%%%%%%%%%%%%%%%%%%%%%%%%%%%%%%%%%%%%%

\subsubsection{Grassmannian formulae}
One of the early signs of positive geometries in the realm of scattering amplitudes came from the realisation that momentum conservation, which is a quadratic constraint in the spinor helicity space\footnote{Spinor helicity variables are introduced in Appendix \ref{app:spaces}.}, can be linearised by introducing auxiliary spaces. More explicitly, the condition $\sum_{i=1}^n \lambda_i^a\widetilde\lambda_i^{\dot a}=0$ can be linearised by introducing an auxiliary $k$-plane in $n$-dimensions, $C=(c_{i}^a)$, such that
\begin{equation}
C^\perp\cdot \lambda=0 \qquad C \cdot \tilde\lambda=0 \,,
\end{equation}
where $C^\perp$ is the orthogonal complement of $C$.

This led to a remarkable development proposed in \cite{ArkaniHamed:2009dn}, where the leading singularities of the $\mathcal{N}=4$ sYM 
N${^{k-2}}$MHV $n$-point amplitudes written in twistor space were described by an integral over the space of $k$-planes in $n$ dimensions, the Grassmannian $G(k,n)$, along suitable closed
contours. Therefore the tree-level amplitudes can be written as
\begin{equation}
A^{\text{\tiny tree}}_{n,k} = \int_\gamma \frac{\prod_{a,i} dc^a_{i}}{GL(k)(1\ldots k)(2\ldots k+1)\ldots(n \ldots n+k-1)} \prod_{a=1}^k \delta^{4|4}\left(\sum_{i=1}^n c^a_{i} \mathcal{W}_i\right) \,,
\label{ACCK}
\end{equation}
where $\mathcal{W}_i^{\mathcal{A}}$ are the super-twistor variables, see Appendix \ref{app:spaces}, and $\gamma$ is a closed contour.
The denominator consists of the cyclic product of the minors $\mathcal{M}_i=(i\, i+1...i+k-1)$, \emph{i.e.}~the determinants of $(k \times k)$ submatrices of the  matrix $C$. 
The contour $\gamma$ can be determined by using \emph{e.g.}~the BCFW recursion relations, and performing the integral \eqref{ACCK} reduces to evaluating a sum of residues, with each residue corresponding to a positroid cell in the positive Grassmannian.

A very similar formula 
was proposed in \cite{Mason:2009qx}  in terms of momentum supertwistors $\mathcal{Z}_i^{\mathcal{A}}$: one can rewrite the amplitude as 
\begin{equation}
A^{\text{\tiny tree}}_{n,k}=A^{\text{\tiny tree}}_{n,2}\,W_{n,k'}\,,
\label{AandW}
\end{equation}
where $W_{n,k'}$ is the tree-level expectation value of the polygonal light-like Wilson loop dual to the amplitude, and we introduced $k'=k-2$. Then $W_{n,k'}$ can be evaluated from 
\begin{equation}
W_{n,k'} = \int \frac{\prod_{a,i} dt^a_{i}}{GL(k')(1\ldots k')(2\ldots k'+1)\ldots(n \ldots n+k'-1)} \prod_{a=1}^{k'} \delta^{4|4}\left(\sum_{i=1}^n t^a_{i} \mathcal{Z}_i\right) \,.
\label{MS}
\end{equation}

The residues of the Grassmannian integral are in one-to-one correspondence with individual BCFW diagrams. Moreover, the BCFW recursion relations can be solved in various independent ways and the identity between results can be understood as a consequence of the residue theorem for these integrals.  
This observation led Hodges \cite{Hodges:2009hk} to argue that the NMHV tree-level amplitude can be thought of as the volume of a particular polytope in momentum twistor space, for which the different BCFW solutions represent different triangulations.
This idea motivated the search for a geometric representation of amplitudes and culminated with the formulation of the amplituhedron \cite{Arkani-Hamed:2013jha}, which we will describe in the next section.

%%%%%%%%%%%%%%%%%%%%%%%%%%%%%%%%%%%%%%%%%%%%%%%%%%%%%%%%%

\subsubsection{Symmetries of scattering amplitudes}

An important property of $\mathcal{N} = 4$ sYM in the planar limit is the fact that it possesses a Yangian symmetry, which is an algebraic manifestation of its quantum integrability.
Indeed, the Lagrangian of $\mathcal{N} = 4$ sYM is invariant under the superconformal group PSU$(2,2|4)$. Moreover, in the planar limit a hidden symmetry not visible at the Lagrangian level appears: the dual superconformal symmetry. This is a second, distinct copy of PSU$(2,2|4)$. 
The combination of the two superconformal symmetry algebras forms a Yangian structure, whose definition we recall briefly in the following. 
Let us call  $\mathfrak{g}$ the simple Lie algebra generated by the generators $J^{(0)}_a$:
 \begin{equation}
\label{levelzero}
[J^{(0)}_a, J^{(0)}_b] = f_{ab}^{~~c} J^{(0)}_c \,,
\end{equation}
where $f_{ab}^{~~c}$ are the structure constants of $\mathfrak{g}$ and $a = 1,\ldots, \mathrm{dim\, \mathfrak{g}} $. The $J^{(0)}_a$'s form the so-called level-zero Yangian generators. The Yangian $Y(\mathfrak{g})$ of a Lie algebra $\mathfrak{g}$ is the Hopf algebra generated by the set of $J^{(0)}_a$'s together with another set $J^{(1)}_a$, the level one,  which obeys
\begin{equation}
\label{levelone}
[J^{(0)}_a, J^{(1)}_b] = f_{ab}^{~~c} J^{(1)}_c \,,
\end{equation}
and therefore transforms in the adjoint representation of $\mathfrak{g}$.

Since $\mathcal{N}=4$ sYM  is a superconformal field theory, one expects this to be reflected in the structure of its scattering amplitudes. This turns out to be true for tree-level amplitudes  but not at loop level, where the presence of infrared divergences breaks the symmetry.
If we denote with $j_a$ any generator of the superconformal algebra $\mathfrak{psu}(2,2|4)$
we can write\footnote{In fact \eqref{scs} is not completely exact, because of the so-called holomorphic anomaly.}
\begin{equation}
j_a \mathcal{A}^{\mathrm{tree}}_n = 0. 
\label{scs}
\end{equation}
At loop level, the infrared effects can be taken into account by deforming the superconformal generators: as for example in \cite{Bargheer:2009qu,Sever:2009aa,Beisert:2010gn} where it was shown how to redefine them to restore the symmetry at one loop.
The dual superconformal symmetry is generated by a set of $J_a$'s being the dual copy of $\mathfrak{psu}(2,2|4)$. Through a suitable modification of (some of) the dual superconformal generators,
one can show the invariance of $\mathcal{A}_n$ at tree level
\begin{equation}
j'_a \mathcal{A}^{\mathrm{tree}}_n = 0 \,.
\end{equation}
In \cite{Drummond:2009fd} it was shown that the generators $j_a$ \eqref{scs} together with one $j'_a$ generate the Yangian of the superconformal algebra, $Y(\mathfrak{psu}(2,2|4))$. 
If instead we consider the amplitude with the $\mathrm{MHV}$-part factorised out, \emph{i.e.}~$W_{n,k}$ in \eqref{AandW},
the dual superconformal generators are the level zero, and one (suitable modified) superconformal generator forms the level-one.
For details see \emph{e.g.}~the review \cite{Ferro:2018ygf} and references therein.

While proving the Yangian invariance in the spinor helicity and dual spaces is rather difficult, and was explicitly verified only on a limited number of cases, using formulae \eqref{ACCK} and \eqref{MS} allows one to beautifully check it for any $n$ and $k$ \cite{Drummond:2010qh}. These formulae are invariant under the Yangian $Y\big(\mathfrak{psl}(4|4)\big)$, which in momentum twistor space is generated by 
\begin{align}\label{yangianampl}
(J^{(0)})^\mathcal{A}_{\;\mathcal{B}}&=\sum_{i=1}^n \mathcal{Z}_i^\mathcal{A}\frac{\partial}{\partial \mathcal{Z}^\mathcal{B}_i} \,,%\nonumber \\
\qquad (J^{(1)})^\mathcal{A}_{\;\mathcal{B}}=\sum_{i<j} \left(\mathcal{Z}_i^\mathcal{A}\frac{\partial}{\partial \mathcal{Z}^\mathcal{C}_i}\mathcal{Z}_j^\mathcal{C}\frac{\partial}{\partial \mathcal{Z}^\mathcal{B}_j}-(i\leftrightarrow j)\right) \,.
\end{align}
A similar set of generators can be written in the twistor space.

%%%%%%%%%%%%%%%%%%%%%%%%%%%%%%%%%%%%%%%%%%%%%%%
%%%%%%%%%%%%%%%%%%%%%%%%%%%%%%%%%%%%%%%%%%%%%%%
%%%%%%%%%%%%%%%%%%%%%%%%%%%%%%%%%%%%%%%%%%%%%%%

\subsection{Amplituhedron} 
\label{sec:amplituhedron}
We start our journey through positive geometries relevant for physics with two prime examples, the tree amplituhedron and the loop amplituhedron. The tree amplituhedron $\mathcal{A}_{n,k'}^{(m)}$ is a positive geometry encoding the tree-level scattering amplitudes in the momentum super-twistor space\footnote{For this reason the amplituhedron describes the dual polygon Wilson loop, which suggests that it should rather be called Wilsonahedron.} and we can think of it as the generalisation of projective polytopes into the Grassmannian space, in the same way as the positive Grassmannian is the generalisation of a projective simplex. Originally, the tree amplituhedron was defined in \cite{Arkani-Hamed:2013jha} using an auxiliary Grassmannian space; 
it can be however translated directly to  momentum twistor space by performing a particular projection. In both spaces the points inside the tree amplituhedron satisfy particular positivity conditions \cite{Arkani-Hamed:2017vfh}, which uniquely determine them. Finally, using these positivity conditions, the tree amplituhedron can be defined directly in the momentum twistor space as the intersection of a subset of points satisfying particular sign patterns with an affine subspace, without any reference to an auxiliary space.
The majority of known results for tree amplituhedra has been found for $m=1$, $m=2$ and for the physical case $m=4$, and we will mostly focus on these cases. In particular, the complete boundary structure of these spaces is known for $m=1,2$ and an explicit form of canonical forms can be found in all three cases (although there is no closed formula for $m=4$).

The loop amplituhedron $\mathcal{A}_{n,k'}^{\ell\text{\tiny -loop}}$ has been defined in \cite{Arkani-Hamed:2013jha} and provides a positive geometry for integrands of loop amplitudes in $\mathcal{N}=4$ sYM. 
It is defined as the image of a space, generalizing the positive Grassmannian, through a linear map. This construction is available only for the case relevant for physics $m=4$.
The loop amplituhedron also satisfies particular sign patterns \cite{Arkani-Hamed:2017vfh}. While for one-loop amplitudes and for the four-point $\text{MHV}$ amplitude at any loop various results are available, at the moment not much is known beyond these cases.

%%%%%%%%%%%%%%%%%%%%%%%%%%%%%%%%%%%%%%%%%%%%%%%%%%%%%

\subsubsection{Tree Amplituhedron}

%%%%%%%%%%%%%%%%%%%%%%%%%%%%%%%%%%%%%%%%%%%%%%%%%%%%%

\paragraph{Original definition.}
Let us consider a positive matrix $Z\in M_+(m+k',n)$ with entries $Z^A_{i}$ for $A=1,\ldots m+k'$, $i=1\ldots,n$. These will be later reinterpreted as the bosonisation of the momentum twistors \eqref{eq:momtw}.
The tree amplituhedron $\mathcal{A}_{n,k'}^{(m)}$ is defined as the image of the map
\begin{equation}\label{PhiZ}
\Phi_Z:G_{+}(k',n)\to G(k',m+k')\,,
\end{equation}
which to each element $C\in G_+(k',n)$, where $C=(c_{\alpha}^{\,\,i})$, associates $Y=\Phi_Z(C)=c\cdot Z$, or in components
\begin{equation}  
\label{eq:YcZ}
Y^A_\alpha = \sum_{i=1}^n c_{\alpha}^{\,\, i}Z^A_i\,.
\end{equation}
On $\mathcal{A}_{n,k'}^{(m)}\subset G(k',m+k')$ one can define a $(k'\cdot m)$-dimensional canonical differential form $\mathbf\Omega_{n,k'}^{(m)}$, the  \emph{volume form},  with logarithmic singularities on all boundaries of the space:
\begin{equation} 
\label{Omeganotilde}
\mathbf\Omega_{n,k'}^{(m)}(Y,Z)=\prod_{\alpha=1}^{k'} \< Y_1\cdots Y_{k'} \,d^m Y_{\alpha} \> \; \Omega_{n,k'}^{(m)}(Y,Z) \,,
\end{equation}
where $\Omega_{n,k'}^{(m)}$ is the \emph{volume function} and $\prod_{\alpha=1}^{k'} \< Y_1\cdots Y_k' \,d^m Y_{\alpha} \>$ the standard measure on the Grassmannian $G(k',m+k')$.
 We will describe more extensively the volume form later on. The geometric space $\mathcal{A}_{n,k'}^{(m)}$ together with the form $\mathbf{\Omega}_{n,k'}^{(m)}$ is (conjecturally) a positive geometry for all $n,k'$ and $m$.
 
Interestingly, the tree amplituhedron $\mathcal{A}_{n,k'}^{(m)}$ recovers familiar objects for special values of its labels: if $Z$ is a square matrix, \emph{i.e.}~$m+k' = n$, then $\mathcal{A}_{n,k'}^{(n-k')}$ is isomorphic to the positive Grassmannian $G_+(k',n)$. If $k' = 1$, then  $\mathcal{A}_{n,1}^{(m)}$ is a projective cyclic polytope \cite{Sturmfels}. Finally, when $m = 1$, $\mathcal{A}_{n,k'}^{(1)}$ can be identified with the complex of bounded faces of a cyclic hyperplane arrangement \cite{Karp:2016uax}. Most importantly, the canonical form $\mathbf{\Omega}_{n,k'}^{(4)}$ encodes the tree-level amplitude $A^{\text{\tiny tree}}_{n,k'+2}$.

%%%%%%%%%%%%%%%%%%%%%%%%%%%%%%%%%%%%%%%%%%%%%%%

\paragraph{Topological description.} 
The amplituhedron definition implies that the points inside the amplituhedron  satisfy certain sign patterns \cite{Arkani-Hamed:2017vfh}. In particular, it is straightforward to show 
that if $Y\in \mathcal{A}_{n,k'}^{(m)}$ then
\begin{equation}\label{eq:signseven}
\langle Y(i_1 i_{1}+1) \ldots (i_{\frac{m}{2}} i_{\frac{m}{2}}+1 ) \rangle > 0  \,,
\end{equation}  
when $m$ is even and
\begin{equation}
(-1)^k \langle Y 1(i_1 i_{1} + 1) \ldots(i_{\frac{m-1}{2}} i_{\frac{m-1}{2}}+1)\rangle>0,\qquad  \langle  Y (i_1 i_1 + 1) \ldots(i_{\frac{m-1}{2}} i_{\frac{m-1}{2}}+1)n \rangle > 0\,,
\end{equation}
when $m$ is odd. Moreover, the following sequence of brackets
\begin{equation}
\label{eq:sequence}
\{\langle Y12 \ldots (m-1)m\rangle,\ldots , \langle Y 12 \ldots (m-1)n\rangle \}\,
\end{equation}
has exactly $k'$ sign flips. It was argued in \cite{Arkani-Hamed:2017vfh} that also the converse is true and we can define the amplituhedron by demanding these sign patterns. This will allow us in the following to introduce a definition of the amplituhedron which does not refer to any auxiliary space. 

%%%%%%%%%%%%%%%%%%%%%%%%%%%%%%%%%%%%%%%%%%%%%%%

\paragraph{Definition in the kinematic space.}

From the point of view of scattering amplitudes, the natural space is the physical kinematic space of $z$'s, see \eqref{eq:momtw}, while the $Y$-space on which the amplituhedron is defined plays the role of an auxiliary space. 
In order to define the amplituhedron directly on the kinematic space, let us first notice that each element $Y\in G(k',m+k')$ defines an $m$-dimensional subspace in $n$ dimensions in the following way: let $Y^\perp$ be an orthogonal complement of $Y$ and let us define 
\begin{equation}\label{small.Z}
z^a_i=(Y^\perp)_A^a\,Z_i^A \,.
\end{equation}
Formula \eqref{small.Z} provides a map $\Xi:G(k',m+k')\to \mathcal{Z}(n)$ from the auxiliary space $G(k',m+k')$, where the amplituhedron lives, to the kinematic space $\mathcal{Z}(n)$ whose elements are the bosonic components of momentum supertwistors, namely $\mathcal{Z}(n)=(z)$.
By composing this map with $\Phi_{Z}$ we can define the amplituhedron $\mathcal{A}_{n,k'}^{(m),z}$ directly in the momentum twistor space as the image of the positive Grassmannian $G_+(k',n)$:
\begin{equation}
\mathcal{A}^{(m),z}_{n,k'}= \Xi\left(\Phi_{Z}\left(G_+(k',n)\right)\right)\,.
\end{equation}
The canonical form on this space $\mathcal{A}^{(m),z}_{n,k'}$ depending on the $z$'s variables can be found by using the push-forward $\Xi_*$ from $\mathcal{A}^{(m)}_{n,k'}$:
\begin{equation}
\mathbf{\Omega}_{n,k'}^{(m),z}=\Xi_*\,\mathbf{\Omega}_{n,k'}^{(m)} \,.
\end{equation} 
This opens up the possibility of a description of the amplituhedron $\mathcal{A}_{n,k'}^{(m)}$ without the need of introducing auxiliary variables $Y$. 
To show this, let us start by writing $Y$ in a particular patch of the Grassmannian space as
\begin{equation}
  Y^A_\alpha= 
\begin{pmatrix} 
- y_\alpha^a\\
1_{k'\times k'}
\end{pmatrix} \quad \rightarrow \quad
  ( Y^\perp)^{ a}_{ A} = 
\left(\begin{tabular} {c|c}
$1_{m\times m}$&
$ y^{ a}_{ \alpha}$
\end{tabular} \right) \,.
\end{equation}
Then, by decomposing the matrix $Z$ in the following way
\begin{equation}\label{big.Z}
Z_i^A = 
\begin{pmatrix} 
z_i^{* a} \\
\Delta^\alpha_i
\end{pmatrix} ,
\end{equation}
where $(z^*)$ is a fixed $m$-plane in $n$ dimensions and $\Delta$ is a fixed $k'$-plane in $n$ dimensions, 
we have that \eqref{small.Z} can be directly written as
\begin{equation}\label{subspace.Z}
 z_i^a = z^{*a}_i+y_{\alpha}^a \Delta_i^\alpha \,.
\end{equation} 
This allows us to define the following $(k'\cdot m)$-dimensional subspace of the kinematic space 
\begin{equation}\label{subspace.definition}
\mathcal{V}^{(m)}_{n,k'} 
=
 \{z_i^a:
     z_i^a = z^{*a}_i+y_{\alpha}^a \Delta_i^\alpha 
\} \,,
\end{equation}
where we assume that when we assemble $z^*$ and $\Delta$ as in \eqref{big.Z} then $Z$ is a positive matrix. 
We also define a winding space $\mathcal{W}^{(m)}_{n,k'}$ as the subset of  kinematic space consisting of points satisfying conditions \eqref{eq:signseven}-\eqref{eq:sequence} after we project them down to the kinematic space (which results in removing $Y$ from the brackets). For example, for $m=2$ this winding space takes the following form: 
\begin{eqnarray}
\mathcal{W}^{(2)}_{n,k'}=&\{ \langle i i+1\rangle_z >0 \,\text{and the sequence} \left\{ \langle 12\rangle_z,\langle 13\rangle_z,\ldots ,\langle 1n\rangle_z \right\} \text{has $k'$ sign flips}\} \,,
\end{eqnarray}
where we have defined the brackets $\langle ij\rangle_z:=z_i^1 z_j^2-z_i^2 z_j^1$.
The amplituhedron $\mathcal{A}_{n,k'}^{(m),z}$ can then be alternatively defined directly in kinematic space as the intersection:
$$
\mathcal{A}_{n,k'}^{(m),z}=\mathcal{V}^{(m)}_{n,k'}\cap \mathcal{W}^{(m)}_{n,k'}\,.
$$

%%%%%%%%%%%%%%%%%%%%%%%%%%%%%%%%%%%%%%%%%%%%%%%%%%%%%%%%%%%%%

\subsubsection{Boundaries and Volume form}

The amplituhedron $\mathcal{A}_{n,k'}^{(m)}$ is (conjecturally) a positive geometry: its canonical form $\mathbf{\Omega}_{n,k'}^{(m)}$ has logarithmic singularities on all its boundaries.
The first step to rigorously check this statement is to find the boundary stratification of $\mathcal{A}_{n,k'}^{(m)}$. The general structure of the amplituhedron boundaries is however unknown for $m>2$. Despite this fact, it is often enough to know the facets of the amplituhedron, \emph{i.e.}~the co-dimension one boundaries, to find a candidate logarithmic form.  
The facets of the amplituhedron 
are known for the first few values of $m$:
\begin{itemize}
\item $m=1$: $\langle Y i \rangle = 0$, for $i=1,\ldots,n$\,,
\item $m=2$: $\langle Yii+1\rangle=0$, for $i=1,\ldots,n$\,,
\item $m=4$: $\langle Yii+1jj+1\rangle=0$,	 for $i < j =1,\ldots,n$ \,.
\end{itemize}
Beyond facets, we know the complete boundary stratification only for $m=1,2$: for $m=1$ \cite{Karp:2016uax} the amplituhedron $\mathcal{A}_{n,k'}^{(1)}$ can be identified with the well-known complex of bounded faces of a cyclic hyperplane arrangement, while for $m=2$ the complete boundary stratification of $\mathcal{A}_{n,k'}^{(2)}$ has been studied in \cite{Lukowski:2019kqi}. At the moment, the boundary stratification for the most interesting, physical case $m=4$ is not known.

Knowing the boundaries of the amplituhedron we are now looking for a differential form with logarithmic singularities on these boundaries. As we have already advertised there are various different methods to find such canonical form, as we describe below.

%%%%%%%%%%%%%%%%%%%%%%%%%%%%%%%%%%%%%%%%%%%%%%%

\paragraph{Triangulations.}

The dimension of the positive Grassmannian $G_{+}(k',n)$ is 
larger than the dimension of the amplituhedron $\mathcal{A}_{n,k'}^{(m)}$. This means that the map $\Phi_Z$ is not injective and the image is covered infinitely many times. One way to find the canonical form is to find a triangulation of the amplituhedron, namely a collection of positroid cells $\mathcal{S}=\{S_{\sigma} \}$ in $G_{+}(k',n)$ with each mapping injectively to its image and their images being disjoint and dense in the amplituhedron. Since we know canonical forms $\omega_\sigma$ for each cell in $\mathcal{S}$, then the volume form $\mathbf{\Omega}_{n,k'}^{(m)}$ can be found by evaluating the push-forward of the canonical forms $\omega_\sigma$ via the function $\Phi_Z$ and then summing over all positroid cells in the triangulation $\mathcal{S}$
\begin{equation}
\mathbf{\Omega}_{n,k'}^{(m)}=\sum_{S_\sigma\in \mathcal{S}}(\Phi_Z)_*\,\omega_\sigma \,.
\end{equation}
The result of the push-forward is a logarithmic differential form on $G(k',m+k')$ which can be written as
\begin{equation}\label{Omega.dlog}
\mathbf{\Omega}_{n,k'}^{(m)}=\sum_{S_\sigma\in\mathcal{S}} \mathrm{d}_Y\mathrm{log}\,\alpha^\sigma_1(Y,Z)\wedge \mathrm{d}_Y\mathrm{log}\,\alpha^\sigma_2(Y,Z)\wedge \ldots\wedge \mathrm{d}_Y\mathrm{log}\,\alpha^\sigma_{k m}(Y,Z)\,,
\end{equation}
where $\alpha^\sigma_i(Y,Z)$ are the canonical positive coordinates parametrizing the cell $S_\sigma$.

Triangulations of amplituhedra have been studied for various values of $m$. For $m=4$ a large class of triangulations can be found from BCFW recursion relations. For $m=2$ they were studied in \cite{Karp:2017ouj}, where the number of triangles in each triangulation was conjectured to be a Narajama number, while in \cite{Bao:2019bfe} it was rigorously proved that $\mathcal{A}_{n,k'}^{(2)}$ admits a triangulation and in \cite{Lukowski:2020dpn} its triangulations 
were related to positroidal triangulations of the hypersimplex $\Delta_{k'+1,n}$. For $m=1$ examples of BCFW-like triangulations have been studied in \cite{Karp:2016uax}.
In the following we summarise the known results for the canonical forms of amplituhedra coming from triangulations: 
\begin{itemize}
\item $m=1$ case \cite{Arkani-Hamed:2017tmz}: for even $k'$ we have
\begin{equation}
\mathbf\Omega_{n,k'}^{(1)}=\prod_{\alpha=1}^{k'} \< Y_1\cdots Y_{k'} \,d Y_{\alpha} \> 
\sum_{2\leq j_1-1<j_1<\cdots<j_\frac{k'}{2}-1< n}[1,j_1-1, j_1,\ldots, j_\frac{k'}{2}-1, j_\frac{k'}{2}]\,,
\end{equation}
while for odd $k'$
\begin{equation}
\mathbf\Omega_{n,k'}^{(1)}=\prod_{\alpha=1}^{k'} \< Y_1\cdots Y_{k'} \,d Y_{\alpha} \> 
\sum_{2\leq j_1-1<j_1<\cdots<j_\frac{k'-1}{2}-1< n-1}[1,j_1-1, j_1,\ldots, j_\frac{k'-1}{2}-1, j_\frac{k'-1}{2}]\,,
\end{equation}
where
\begin{equation}
[j_0, j_1,\ldots , j_{k'}] := \frac{\langle j_0 \ldots j_{k'}\rangle}{ 
\langle Y j_0\rangle \cdots \langle Y j_{k'}\rangle}\,,
\end{equation}
\item $m=2$ case: see  \cite{Arkani-Hamed:2017tmz} or \cite{Lukowski:2019sxw}:
\begin{equation}
\mathbf\Omega_{n,k'}^{(2)}=\prod_{\alpha=1}^{k'} \< Y_1\cdots Y_{k'} \,d^2 Y_{\alpha} \> \sum_{2\leq i_1<\cdots i_{k'}\leq n-1}[1, i_1, i_1+1;\ldots ; 1, i_{k'}, i_{k'}+1] \,,
\end{equation}
where
\begin{equation}
[p_1, q_1, r_1;\ldots ; p_{k'}, q_{k'}, r_{k'}] =
\frac{[\langle (Y^{k'-1})^{s_1} p_1q_1r_1
\rangle\ldots \langle (Y^{k'-1})^{s_{k'}} p_{k'}q_{k'}r_{k'}\rangle
\epsilon_{s_1\cdots s_{k'}}]^{k'}}{2^{k'} \langle Y p_1q_1\rangle \langle Y q_1r_1\rangle \langle Y p_1r_1\rangle \cdots \langle Y p_{k'}q_{k'}\rangle \langle Y q_{k'}r_{k'}\rangle \langle Y p_{k'}r_{k'}\rangle}\,,
\end{equation}
\item $m=4$ case: the explicit answer for all $n$ is known only for $k'=1$ for which the amplituhedron is a cyclic polytope, see \emph{e.g.}~\cite{Arkani-Hamed:2017tmz},
\begin{equation}
\mathbf\Omega_{n,1}^{(4)}= \< Y_1 \,d^4 Y_{1} \>\sum_{i<j}[1ii+jj+1]\,,\end{equation}
where
\begin{equation}
[i_1i_2i_3i_4i_5]=\frac{\langle i_1 i_2i_3i_4i_5\rangle^4}{\langle Yi_1i_2i_3i_4\rangle\langle Yi_2i_3i_4i_5\rangle\langle Yi_3i_4i_5i_1\rangle\langle Yi_4i_5i_1i_2\rangle\langle Yi_5i_1i_2i_3\rangle}\,
\end{equation}
is a bosonised version of R-invariants. Beyond $k'=1$, the BCFW triangulation as a sum over positroid cells can easily be found, and is for example implemented in the Mathematica package \texttt{positroid} \cite{Bourjaily:2012gy}, however there is no known explicit general answer in this case.
\end{itemize}

An interesting problem is to classify all possible triangulations, which produce a large set of possible representations of canonical forms and therefore of amplitudes. This problem has been studied for $m=2$ in \cite{Lukowski:2020dpn}, where a subclass of triangulations, called regular triangulations, has been identified with the finest cones in the positive tropical Grassmannian \cite{troppos}. Knowing all regular triangulations one can define a secondary polytope, each vertex of which is a regular triangulation. For convex $n$-gons, which are the amplituhedra $\mathcal{A}_{n,1}^{(2)}$, the secondary polytope is the associahedron. For $m=2$ and general $k'$ the secondary polytope is given by the dual of the positive tropical fan \cite{Lukowski:2020dpn}. For $m=4$ a construction of the secondary geometry is still unknown. One possible approach to find this geometry is to generalise the Jeffrey-Kirwan construction of amplituhedron volume forms \cite{Ferro:2018vpf} beyond $k'=1$.

%%%%%%%%%%%%%%%%%%%%%%%%%%%%%%%%%%%%%%%%%%%%%%%

\paragraph{Contour integrals.}

An alternative way to compute the volume function is given by evaluating the following integral
\begin{equation}\label{volume.function.integral}
\Omega_{n,k'}^{(m)}=\int_\gamma \frac{\mathrm{d}^{k'\cdot n}c_{\alpha i}}{(12\ldots k')(23\ldots k'+1)\ldots (n1\ldots k'-1)}\prod_{\alpha=1}^{k'}\delta^{m+k'}(Y^A_\alpha-\sum_i c_{\alpha i}Z^A_i)\,,
\end{equation}
taken over a suitable closed contour $\gamma$, in analogy with the Grassmannian integral \eqref{MS}. The contour can be determined for example by  using the BCFW recursion relations and it selects a particular combination of poles of the integrand. Each residue corresponds to the volume function on a ``triangle'' in the tree amplituhedron. Then the volume function  $\Omega_{n,k'}^{(m)}$ is calculated as a particular sum of such residues. 
There have been few attempts to fix the contour of integration without making reference to \emph{e.g.}~BCFW recursion relations. One can use the ``Feynman prescription" and modify the denominators of \eqref{volume.function.integral} by adding a positive $i\epsilon$ to each factor in the denominator \cite{Ferro:2015grk,Arkani-Hamed:2017tmz}. Then, after solving the delta function in \eqref{volume.function.integral}, one ends up with a $k'\times(n-m-k')$-dimensional integral which can be performed over the product of real lines. Using the positivity of external data one can show that this contour produces the correct answer for $k'=1$ and even $m$. A generalisation to any $k'$ is not known at the moment. Alternatively, for $k'=1$ or $k'=n-m-1$, the contour can be fixed using the Jeffrey-Kirwan prescription \cite{Ferro:2018vpf}.

%%%%%%%%%%%%%%%%%%%%%%%%%%%%%%%%%%%%%%%%%%%%%%%

\paragraph{Poles and Zeros.}

 An alternative approach was suggested in \cite{Arkani-Hamed:2014dca}, where canonical forms were found by demanding their regularity everywhere outside of $\mathcal{A}_{n,k'}^{(m)}$. 
This was based on the observation that  
only a small subset of intersections of the co-dimension one boundaries are themselves the amplituhedron boundaries. The majority of intersections is located outside the amplituhedron and the canonical form must be regular when approaching them. 

Let us present an example for $m=2$. The facets of $\mathcal{A}_{n,k'}^{(2)}$ are characterised by  $\langle Y i i+1\rangle = 0$ and  positivity implies $\langle Y i i+1\rangle >0$ for all points $Y$ inside the amplituhedron. Therefore a factor of $\langle Y i i+1\rangle$ for any $i$  has to appear in the denominator of $\Omega_{n,k'}^{(2)}$:
\begin{equation}
\Omega_{n,k'}^{(2)} = \prod_{j=1}^{k'} \langle Y d^2Y_j\rangle \frac{\mathcal{N}(Y)}{\langle Y12 \rangle \langle Y23\rangle \ldots \langle Yn1\rangle} \,.
\end{equation}
By taking residues of $\Omega_{n,k'}^{(2)} $ we can access the lower-dimensional boundaries of $\mathcal{A}_{n,k'}^{(2)}$. The residues are found by setting $\langle Y \ldots \rangle = \langle Y \ldots \rangle = \cdots =0$ and only small subset of them will correspond to amplituhedron boundaries -- the remaining are spurious poles and the numerator $\mathcal{N}(Y)$ has to vanish when $Y$ approaches them. This requirement is sufficient to determine the numerator uniquely. Moreover, the numerator $\mathcal{N}(Y)$ is always positive for all points inside the amplituhedron, which implies that the differential form is always positive.
This positivity is conjectured to hold true for all $n, k'$ (at all loop orders).
This approach allows one to find new formulas for amplitudes, not involving triangulations, coming directly from the global geometry of the amplituhedron.

%%%%%%%%%%%%%%%%%%%%%%%%%%%%%%%%%%%%%%%%%%%%%%%

\paragraph{Dual geometry.}

Finally, we want to mention that for projective polytopes the canonical function can be found by computing the volume of the dual polytope. 
In the case $k'=1$ and any $m$, the amplituhedron $\mathcal{A}_{n,1}^{(m)}$ is a cyclic polytope in $\mathbb{P}^{m}$ and we can rewrite its canonical form $\mathbf{\Omega}_{n,1}^{(m)}$ as
\begin{equation}
\mathbf{\Omega}^{(m)}_{n,1} = \langle Y d^m Y\rangle \int_{\tilde{\mathcal{A}}} \frac{\langle W d^m W\rangle}{(W\cdot Y)^{m+1}} \,,
\end{equation}
where the dual polytope is defined as
\begin{equation}
\tilde{\mathcal{A}}_{n,1}^{(m)}=\left\{W\in \mathbb{P}^m: W\cdot Y\geq 0 \text{ for all } Y\in \mathcal{A}_{n,1}^{(m)}\right\}\,,
\end{equation}
and the integral computes its projective volume.
For $k'>1$ there is no known generalisation of the corresponding dual geometry yet.

%%%%%%%%%%%%%%%%%%%%%%%%%%%%%%%%%%%%%%%%%%%%%%%%%%%%%%%%%%%%%

\subsubsection{Amplitudes from Amplituhedron}

We have already mentioned that tree-level scattering amplitudes $A_{n,k}^{\text{\tiny tree}}$ in $\mathcal{N}=4$ sYM can be extracted from canonical forms of the amplituhedron $\mathcal{A}_{n,k-2}^{(4)}$. There are two ways in which we can calculate them:
\begin{itemize}
\item
 Taking the original definition of the amplituhedron in the Grassmannian space, we start from  the volume function $\Omega_{n,k'}^{(4)}$ and localise $Y$ on the reference point $Y^* =$\scalebox{1}{$ ( 0_{4\cdot k'} \; \big| \; \idM_{k'})^{\!T}$}. Furthermore, we parametrise the matrix $Z$ as
\begin{equation}\label{bosonizedZ}
Z_i^A = \left ( 
\begin{tabular}{c}
  $z_i^a$ \\ $ \phi_1^{\mathsf A}\;\chi_{i\mathsf A}$ \\ \vdots\\ $ \phi_k^{\mathsf A}\;\chi_{i\mathsf A}$
  \end{tabular} 
  \right ) \,,\qquad 
  \begin{aligned}
  i &= 1,\dots, n\,, \\ A &= 1,\dots, 4+k'\,, \\ a,\mathsf A &= 1, \ldots, 4\,,
  \end{aligned} \,
\end{equation}
and integrate the volume function over the Grassmann-odd parameters $\phi$:
\begin{equation}\label{amplitoampli}
A_{n,k'}^{\text{\tiny tree}}(\mathcal{Z})= \int \diff^{4\cdot k'}\,\phi \;\; \Omega_{n,k'}^{(4)}(Y^*,Z) \, ,
\end{equation}
where  $\mathcal{Z}_i^{\mathcal{A}}=(z_i^a |\chi_i^{\msf A})\equiv(\lambda_i^\alpha,\tl\mu_i^{\dt\alpha}| \chi_i^{\msf A})$ are momentum supertwistors \cite{Hodges:2009hk}, see appendix \ref{app:spaces}.

\item
Alternatively, one can find the amplitude from the canonical form $\mathbf{\Omega}_{n,k'}^{(m),z}$ defined on the kinematic space. Recall that $\mathbf{\Omega}_{n,k'}^{(m),z}$ is a rank $m\cdot k'$ differential form. We can find the amplitude by replacing the differentials $dz$ with the Grassmann-odd variables parametrizing the on-shell superspace $\eta$:
\begin{equation}
A_{n,k'}^{\text{\tiny tree}}(\mathcal{Z})=\mathbf{\Omega}_{n,k'}^{(m),z}\Big|_{dz^a_i\to\eta^a_i} \,.
\end{equation}

\end{itemize}

%%%%%%%%%%%%%%%%%%%%%%%%%%%%%%%%%%%%%%%%%%%%%%%%%%%%%%%

\subsubsection{Yangian invariance}
We have already commented that tree-level scattering amplitudes are Yangian invariant. The bosonised amplitude encoded in the volume function is however not invariant under a straightforward bosonisation of the Yangian generators. Nevertheless, it was showed in \cite{Ferro:2016zmx} that Yangian invariance is still present, even though in a non-standard way. 
Using the Quantum Inverse Scattering method, it was shown that there exists a matrix of functions closely related to the volume function $\Omega_{n,k'}^{(m)}$ which is invariant under the Yangian of $gl(m+k')$. In particular, if we define
\begin{equation}
(J_Y)^A_{\; B}=\sum_{\alpha=1}^{k'} Y^A_\alpha\frac{\partial}{\partial Y_\alpha^B}+k'\,\delta^A_{\;B} \,,
\end{equation}
then the matrix of functions 
\begin{equation}\label{omega.A.B}
\Omega^A_{\;B}(Y,Z) := (J_Y)^A_{\;B} \, \Omega_{n,k'}^{(m)}(Y,Z) \,,
\end{equation}
is annihilated by the Yangian generators of $Y(gl(m+k'))$.

%%%%%%%%%%%%%%%%%%%%%%%%%%%%%%%%%%%%%%%%%%%%%%%%%%%%%%

\subsubsection{Loop level}

Until now we have discussed the tree amplituhedron $\mathcal{A}_{n,k'}^{(m)}$ which encodes tree-level scattering amplitudes $A^{\text{\tiny tree}}_{n,k'+2}$ when we set $m=4$. The natural next step would be to also find a positive geometry which captures further terms in the perturbative expansion \eqref{eq:perturbation} -- this is not known at the moment. There exists however a geometric construction which computes integrands of amplitudes at loop level \cite{Arkani-Hamed:2013jha}: 
the loop amplituhedron  $\mathcal{A}_{n,k'}^{\ell-\text{\tiny loop}}$. As for the tree amplituhedron, it depends on the number of particles $n$ and the helicity sector $k'$, but has been defined so far only for $m=4$, allowing us to omit this label. It is conjectured that $\mathcal{A}_{n,k'}^{\ell-\text{\tiny loop}}$ is a positive geometry calculating the $\ell$-loop integrands contributing to the scattering amplitude $A_{n,k'+2}$.

Similarly to the tree amplituhedron, the loop amplituhedron is defined as the image of a particular space, generalizing the positive Grassmannian, through a linear map. For a given $n$, $k'$ and $\ell$, we denote by $G(k',n; \ell)$ the space which consists of $k'$-planes $C$ in $n$ dimensions together with $\ell$ two-planes $D^{(\ell)}$, living in the $(n-k')$-dimensional complement of $C$. 
A point in $G(k',n; \ell)$  is represented via a $(k'+2\ell) \times n$  matrix $\mathcal{C}$:
\begin{equation}
\mathcal{C} = \left(
\begin{tabular}{c}
$D^{(l_1)}$ \\
\hdashline
$\vdots$ \\
\hdashline
$D^{(l_\ell)}$ \\
\hdashline
$C$ 
\end{tabular}
\right) \,.
\end{equation}
We denote by $G_+(k',n;\ell)$ the positive part of $G(k',n;\ell)$ which is defined by demanding that all the ordered maximal minors of the matrices
\begin{equation}
\left(
\begin{tabular}{c}

$C$ 
\end{tabular}
\right),\qquad
\left(
\begin{tabular}{c}
$D^{(l_1)}$ \\
\hdashline
$C$ 
\end{tabular}
\right),\qquad
\left(
\begin{tabular}{c}
$D^{(l_1)}$ \\
\hdashline
$D^{(l_2)}$ \\
\hdashline
$C$ 
\end{tabular}
\right),\qquad\ldots\qquad
\left(
\begin{tabular}{c}
$D^{(l_1)}$ \\
\hdashline
$\vdots$ \\
\hdashline
$D^{(l_\ell)}$ \\
\hdashline
$C$ 
\end{tabular}
\right),
\end{equation}
are positive for all $l_1,l_2,\ldots,l_\ell=1,\ldots,\ell$ and $l_i\neq l_j$.
These positivity constraints can be seen as the ``echo" of the standard positivity of a bigger $(k'+2\ell) \times (n+2\ell)$ matrix, of which $\ell$ pairs of adjacent columns  have been removed.
The loop amplituhedron is then the image of $G_+(k',n; \ell)$ through the linear map specified by the external data
 \begin{equation}
 \mathcal{A}_{n,k'}^{\ell-\text{\tiny loop}} = \{ \mathcal{Y} \in G(k',4+k'; \ell) \,  ; \, \mathcal{Y} = \mathcal{C} \cdot Z \,,\, \mathcal{C} \in G_{+}(k',n;\ell), Z \in M_+(4+k',n)  \}\,,
\end{equation}
where $\mathcal{Y}$ is a $k'$-plane $Y$ in $(4+k')$ dimensions, together with $\ell$ two-planes $\mathcal{L}^{(l)}$ living in the four-dimensional orthogonal complement of $Y$:
\begin{equation}
\mathcal{Y} = \left(
\begin{tabular}{c}
$\mathcal{L}^{(1)}$ \\
\hdashline
$\vdots$ \\
\hdashline
$\mathcal{L}^{(\ell)}$ \\
\hdashline
$Y$ 
\end{tabular}
\right), \qquad  \qquad \mathcal{L}_{\beta}^{(l),A} = \sum_i D^{(l),i}_{\beta} Z_{i}^A \,.
\end{equation}
One observes that $ \mathcal{A}_{n,0}^{1-\text{\tiny loop}} $ and $ \mathcal{A}_{n,2}^{(2)} $  are formally identical spaces and hence the one-loop $\mathrm{MHV}$ integrands are related to the canonical forms of the $m=2$ tree amplituhedron.

The definition of $\mathcal{A}_{n,k'}^{\ell-\text{\tiny loop}}$ implies that any point inside the loop amplituhedron satisfies  particular sign patterns \cite{Arkani-Hamed:2017vfh}: in addition to \eqref{eq:signseven}-\eqref{eq:sequence} we have the following conditions for each loop $\mathcal{L}^{(l)}$ 
\begin{align}
&[Y\mathcal{L}^{(l)}ii+1]>0\,, \\
&\{[Y\mathcal{L}^{(l)}12],\ldots,[Y\mathcal{L}^{(l)}1n]\}\text{ has $k'+2$ sign flips}\,,
\end{align}
and for any pair of loops $(\mathcal{L}^{(l_1)},\mathcal{L}^{(l_2)})$ we have
\begin{equation}
 [Y\mathcal{L}^{(l_1)}\mathcal{L}^{(l_2)}]>0\,.
\end{equation}
The sign-flip characterisation of the loop amplituhedron is particularly useful when showing that locality and unitarity follow from positivity at loop level \cite{YelleshpurSrikant:2019meu} and to find new representations of canonical forms \cite{Kojima:2020tjf}. It is also useful for determining the branch points of general amplitudes from the loop amplituhedron using the Landau equations, see \cite{Dennen:2016mdk,Prlina:2017azl,Prlina:2017tvx}.

As for other positive geometries, the next step after defining the space is to understand its boundaries and find the canonical forms with logarithmic singularities on all of them. However, not much is known for the general loop amplituhedron. A comprehensive study of the one-loop case was presented in \cite{Bai:2015qoa}, where a Grassmannian integral formula generalizing \eqref{volume.function.integral} was postulated. As for  tree level, a suitable sum of residues of this new integral allows one to find canonical forms for $\mathcal{A}_{n,k'}^{1-\text{\tiny loop}}$. A two-loop study for MHV amplitudes can be found in \cite{Kojima:2018qzz}. Moreover, the study of a particular class of boundaries for the loop amplituhedron, corresponding to particular cuts of loop integrands, has been initiated in \cite{Arkani-Hamed:2013kca} and expanded to all loop orders in \cite{Arkani-Hamed:2018rsk} and \cite{Langer:2019iuo}. Beyond that, the main focus has been on understanding the simplest possible case: the integrands for the four-point MHV amplitude.

The loop amplituhedron definition simplifies significantly for MHV amplitudes. Indeed in this case $k'=0$ and $\mathcal{C}$ is composed only of  matrices $D^{(l)}$: therefore, the only positivity conditions one needs to consider are between these matrices. The situation simplifies even further when considering $n=4$, where the matrices $D^{(l)}\in G_+(2,4)$ can be parametrised as
\begin{equation}
D^{(l)}=\left(\begin{tabular}{cccc}1&$x_l$&0&$-w_l$\\0&$y_l$&1&$z_l$\end{tabular}\right) \,,
\end{equation}
and we only need to impose 
$ 
\mathrm{det} \left(
  \begin{tabular}{c}
   $D^{(l_1)}$  \\
    $D^{(l_2)}$ 
  \end{tabular}
\right) >0
$, for
all pairs $(l_1,l_2)$. This reduces to the following inequalities
\begin{equation}
x_l,y_l,w_l,z_l>0,\qquad (x_{l_1}-x_{l_2})(z_{l_1}-z_{l_2})+(y_{l_1}-y_{l_2})(w_{l_1}-w_{l_2})>0 \,.
 \end{equation}
The canonical form for this geometry, \emph{i.e.}~the four-point $\mathrm{MHV}$ integrand, has been found up to three loops in \cite{Arkani-Hamed:2013kca}. The boundary stratification of the loop amplituhedron $\mathcal{A}_{4,0}^{\ell-\text{\tiny loop}}$ has been described in \cite{Franco:2014csa} and \cite{Galloni:2016iuj} up to three loops.

%%%%%%%%%%%%%%%%%%%%%%%%%%%%%%%%%%%%%%%%%%%%%%%
%%%%%%%%%%%%%%%%%%%%%%%%%%%%%%%%%%%%%%%%%%%%%%%
%%%%%%%%%%%%%%%%%%%%%%%%%%%%%%%%%%%%%%%%%%%%%%%

\subsection{Correlahedron}

As an intermission, we mention another geometry underlying observables in $\mathcal{N}=4$ sYM which also naturally lives in the momentum twistor space: the correlahedron \cite{Eden:2017fow}. The correlahedron $\mathcal{C}_{n,k'}$ is the geometry encoding the $n$-point stress-tensor correlation function, where $k'$ is associated with the fermionic degree of the correlator's expansion in the analytic superspace \cite{Howe:1995md}. As in the amplituhedron story, one bosonises the Grassmann-odd variables parametrising the analytic superspace. This leads to a purely bosonic space: the external data is encoded in a collection of $n$ 2-planes $X_i\in G(2,n+k'+4)$ corresponding to a point in analytic superspace parametrised by a line in the momentum twistor space. Then the correlahedron is defined as a subset of the Grassmannian space $G(n+k',4+n+k')$ in the following way:
\begin{equation}
\mathcal{C}_{n,k'} = \{Y \in G(n+k', n+4+k'): \langle Y X_i X_j \rangle > 0\} \,,
\end{equation}
where the brackets $\langle \rangle$ are determinants of $(n+4+k')\times (n+4+k')$ matrices. It was conjectured in \cite{Eden:2017fow} that the stress-tensor correlation functions can be extracted from the canonical form of $\mathcal{C}_{n,k'}$.

One interesting connection with the amplituhedron we described in the previous section is that the correlahedron geometry can be projected down to the amplituhedron space by taking light-like limits, \emph{i.e.}~limits where consecutive space-time points become light-like separated. 
In this case, the stress-tensor correlator reduces to the square of the light-like polygonal Wilson loop, and hence the square of the scattering amplitude. The light-like limits enforce the two-planes $X_i$ to intersect in twistor space, which allows one to parametrise them as $X_i =(Z_i,Z_{i+1})$. Moreover, the $p$-point light-like limit is obtained by requiring $Y$ to simultaneously lie on multiple boundaries $\langle Y X_i X_{i+1} \rangle = 0, i = 1,\ldots, p$, of the correlahedron. 
Then the maximal, $n$-point, light-like limit reduces the correlahedron space from $G(n+k', 4+n+k')$ to $G(k', 4+k')$ by particular operations of partial freezing and projecting. Geometrically, the limit reduces  $\mathcal{C}_{n,k'}$ to $(\mathcal{A}_{n,k'})^2$:
\begin{equation}
(\mathcal{A}_{n,k'})^2 = \{Y \in G(k', 4+k'): \langle Y ii+1jj+1 \rangle>0 \}\,.
\end{equation}
 Algebraically, the volume form of the correlahedron $\mathcal{C}_{n,k'}$ becomes the volume form of the square of the amplituhedron $(\mathcal{A}_{n,k'})^2$, which encodes the square of the tree-level N$^{k'}$MHV superamplitude, or rather of the Wilson loop expectation value:
\begin{equation}
W^2_{n,k'} = \sum_{\tilde k=0}^{k'} W_{n,\tilde k} W_{n,k'-\tilde k} \,.
\end{equation}
Alternatively, if one takes a non-maximal light-like limit, \emph{i.e.}~the limit where fewer points are light-like separated, $p<n$, the canonical form reduces to the canonical form of the square of the loop amplituhedron.

%%%%%%%%%%%%%%%%%%%%%%%%%%%%%%%%%%%%%%%%%%%%%%%%%
%%%%%%%%%%%%%%%%%%%%%%%%%%%%%%%%%%%%%%%%%%%%%%%%%
%%%%%%%%%%%%%%%%%%%%%%%%%%%%%%%%%%%%%%%%%%%%%%%%%

\subsection{Momentum amplituhedron}
\label{sec:momentum.amplituhedron}

As we have already described, the amplituhedron $\mathcal{A}_{n,k'}^{(m)}$ is defined in momentum twistor space, which provides natural coordinates for Wilson loops. The fact that this space can be also used for scattering amplitudes follows from the Wilson loop/scattering amplitude duality which is a property of planar $\mathcal{N}=4$ sYM. In particular, momentum twistors encode a fixed ordering of particles, from which they cannot be separated. In order to go beyond the planar limit, we need to use twistors or spinor helicity variables. 
In this section we describe a positive geometry defined directly in the spinor helicity space -- the momentum amplituhedron $\mathcal{M}_{n,k}^{(m)}$ -- introduced in \cite{Damgaard:2019ztj} for the physical case $m=4$ and later generalised to any even $m$ in \cite{Lukowski:2020dpn}. 

Before we proceed to the momentum amplituhedron definition, we emphasise one more crucial difference compared to the amplituhedron construction. In order to be able to bosonise spinor helicity variables we need to abandon the on-shell chiral superspace $(\lambda^a,\widetilde\lambda^{\dot a}|\eta^A)$, $a,\dot a=1,2, A=1,\ldots,4$, and instead rewrite the amplitudes in    
the non-chiral superspace. This can be accomplished  by performing a Fourier transform of two of the four Grassmann variables, which leads to a space parametrised by variables $(\lambda^a,\eta^r \,|\, \widetilde\lambda^{\dot a},\widetilde\eta^{\dot r})$, $r,\dot r=1,2$. 
 In this way,  the $SU(4)$ R-symmetry of $\mathcal{N}=4$ sYM is broken.  Then the new R-symmetry indices $(r,\dot r)$ can be associated with the spinor indices $(a,\dot a)$ and, by  the replacement
\begin{equation}
\label{lambda_eta}
\eta^a \to d\lambda^a\, ,\qquad\qquad \widetilde\eta^{\dot a}\to d\widetilde\lambda^{\dot a}\,,
\end{equation} 
one can write any function on the non-chiral superspace as a differential form on its bosonic part. In particular, the tree-level N$^{k-2}$MHV  scattering amplitudes can be written as differential forms of degree $(2(n-k),2k)$ in $(d\lambda,d\widetilde\lambda)$, see  \cite{He:2018okq}.

%%%%%%%%%%%%%%%%%%%%%%%%%%%%%%%%%%%%%%%%%%%%%%%%%%%%%%

\subsubsection{Definition and topological description}
The momentum amplituhedron $\mathcal{M}_{n,k}^{(m)}$ can be defined using similar steps we followed for the ordinary amplituhedron: after specifying positive external data, we define the momentum amplituhedron as the image of the positive Grassmannian through a linear map depending on this external data.

We start by introducing a pair of matrices $(\Lambda,\widetilde\Lambda)$, which provide a bosonisation of the spinor helicity variables $(\lambda,\tilde\lambda)$:
\begin{equation}
\Lambda=\left(\begin{matrix}
\Lambda_{1}&\Lambda_{2}&\ldots&\Lambda_n
\end{matrix}\right)\in M(n-k+\tfrac{m}{2},n),\qquad \widetilde\Lambda=\left(\begin{matrix}
\widetilde\Lambda_{1}&\widetilde\Lambda_{2}&\ldots&\widetilde\Lambda_n
\end{matrix}\right)\in M(k+\tfrac{m}{2},n)\,.
\end{equation}
We demand this external data to be positive which we define as: the matrix $\tilde\Lambda$ is a positive matrix and $\Lambda$ is a twisted positive matrix\footnote{This condition can also be rewritten as the requirement that $\Lambda^\perp$ is a positive matrix, where $\perp$ indicates the orthogonal complement.}, see \cite{Galashin:2018fri} for definition of the latter.
Then, the momentum amplituhedron $\mathcal{M}^{(m)}_{n,k} $ is the image of the positive Grassmannian $G_+(k,n)$ through the map
\begin{equation}
\label{Phi}
\Phi_{(\Lambda,\widetilde\Lambda)}:G_+(k,n)\to  G\left(n-k,n-k+\tfrac{m}{2}\right)\times G\left(k,k+\tfrac{m}{2}\right)\,,
\end{equation}
which to each element of the positive Grassmannian $C=\{c_{\dot\alpha i}\}\in G_{+}(k,n)$ associates a pair of Grassmannian elements $( Y,\widetilde Y)\in G(n-k,n-k+\tfrac{m}{2})\times G(k,k+\tfrac{m}{2})$ in the following way
\begin{align}
Y^A_\alpha=c^\perp_{\alpha i}\,,\qquad\qquad \widetilde Y^{\dot{A}}_{\dot{\alpha}}=c_{\dot\alpha i}\,\widetilde\Lambda_i^{\dot{A}}\,\Lambda_i^A\,,
\label{Y}
\end{align}
where $C^\perp=\{c^\perp_{\alpha i}\}$ is the orthogonal complement of $C$.

One important non-trivial property of the momentum amplituhedron is the fact that it is $\frac{m}{2}\cdot(n-\tfrac{m}{2})$-dimensional. While the dimension of $G(n-k,n-k+\tfrac{m}{2})\times G(k,k+\tfrac{m}{2})$ is 
\begin{equation}
\dim(G(n-k,n-k+\tfrac{m}{2}))+\dim(G(k,k+\tfrac{m}{2}))=\tfrac{m}{2}(n-k)+\tfrac{m}{2}k=\tfrac{m}{2}\cdot n\,,
\end{equation}
the image of the positive Grassmannian $G_{+}(k,n)$ when mapped using $\Phi_{(\Lambda,\widetilde\Lambda)}$  is lower dimensional. Indeed,  the momentum amplituhedron lives in a co-dimension $\frac{m^2}{4}$-surface inside $ G(n-k,n-k+\tfrac{m}{2})\times G(k,k+\tfrac{m}{2})$ satisfying:
\begin{equation}\label{momentum.cons}
P^{a\dot a}=\sum_{i=1}^n\left(Y^\perp \cdot \Lambda\right)^a_i \left(\widetilde Y^\perp \cdot \widetilde\Lambda \right)^{\dot a}_i=0\,.
\end{equation} 
In particular, for $m=4$, one can think about the condition \eqref{momentum.cons} as being equivalent to the four-dimensional momentum conservation written directly in the momentum amplituhedron space. If we project $\Lambda$ and $\tilde\Lambda$ through a fixed $Y$ and $\widetilde Y$, as we will see later, then we find
\begin{equation}
\label{projectY}
\left(Y^\perp \cdot \Lambda\right)^a_i  \to \lambda_i^a\,,\qquad\qquad \left(\widetilde Y^\perp \cdot \widetilde\Lambda \right)^{\dot a}_i\to\widetilde\lambda^{\dot a}_i\,,
\end{equation} 
and the condition \eqref{momentum.cons} reduces to the usual momentum conservation.

As for the amplituhedron, the definition of the momentum amplituhedron implies particular sign patterns, which for $m=4$ were postulated in \cite{He:2018okq}. 
Indeed, one can show that for $(Y,\tilde Y)\in \mathcal{M}_{n,k}^{(4)}$ we have $\langle Yii+1\rangle>0$ and $[ \widetilde Yii+1]>0$. Moreover, the number of sign flips in the sequence
\begin{equation}\label{Y.flips}
\{\langle Y12\rangle,\langle Y13\rangle,\ldots,\langle Y1n\rangle\}\,
\end{equation}
equals $k-2$ and there are $k$ sign flips in the sequence
\begin{equation}\label{Ytilde.flips}
\{[ \widetilde Y12],[\widetilde Y13],\ldots,[ \widetilde Y1n]\}\,.
\end{equation}
Here we introduced the brackets $\langle \rangle$ and $[]$ which are defined as Pl{\"u}cker variables of the matrices $(Y_1,\ldots,Y_{n-k},\Lambda_1,\ldots,\Lambda_n)$ and $(\widetilde Y_1,\ldots,\widetilde Y_{k},\widetilde\Lambda_1,\ldots,\widetilde\Lambda_n)$, respectively. 

A similar sign pattern can also be found beyond $m = 4$. For example for $m=2$, the momentum amplituhedron definition implies that the sign patterns are:
\begin{align}
&\{\langle Y1\rangle,\langle Y2\rangle,\ldots,\langle Yn\rangle\}  \text{ has $k-1$ sign flips} \,, \\
&\{[ \widetilde Y1],[\widetilde Y2],\ldots,[ \widetilde Yn] \}\text{ has $k$ sign flips}  \,.
\end{align}

%%%%%%%%%%%%%%%%%%%%%%%%%%%%%%%%%%%%%%%%%%%%%%%

\subsubsection{Definition in the kinematic space}

The definition \eqref{Y} of the momentum amplituhedron demands the introduction of auxiliary Grassmannian spaces. Here, we want to reproduce the argument we used for the amplituhedron and provide a description of the momentum amplituhedron without reference to these auxiliary spaces, by defining it directly in terms of kinematic data in spinor helicity space. In order to do so, we restrict to the physical case $m=4$ and notice that each element $(Y,\widetilde Y)\in G(n-k,n-k+2)\times G(k,k+2)$ defines a pair of two-dimensional subspaces in $n$ dimensions in the following way: let $Y^\perp$ and $\widetilde Y^\perp$ be orthogonal complements of $Y$ and $\widetilde Y$, respectively. Then we define 
\begin{equation}\label{small.lambda}
\lambda^a_i=(Y^\perp)_A^a\,\Lambda_i^A\qquad\qquad \widetilde\lambda_i^{\dot a}=(\widetilde Y^\perp)_{\dot A}^{\dot a}\,\widetilde\Lambda_i^{\dot A} \,.
\end{equation}
These formulae provide a map from the auxiliary space where the momentum amplituhedron lives to the kinematic space $\mathcal{L}(n)=(\lambda,\widetilde\lambda)$:
\begin{equation}
\Xi: G(n-k,n-k+2)\times G(k,k+2)\to \mathcal{L}(n)\,.
\end{equation}
Composing this map with $\Phi_{(\Lambda,\widetilde\Lambda)}$ we define the momentum amplituhedron directly in the spinor helicity space as the image of the positive Grassmannian $G_+(k,n)$:
\begin{equation}
\mathcal{M}^{(\lambda,\widetilde\lambda)}_{n,k}= \Xi\left(\Phi_{(\Lambda,\widetilde\Lambda)}\left(G_+(k,n)\right)\right) \,.
\end{equation}
The canonical form on the space $\mathcal{M}^{(\lambda,\widetilde\lambda)}_{n,k}$ can be found by using the push-forward $\Xi_*$:
\begin{equation}
\label{omegaL}
\mathbf{\Omega}^{(\lambda,\tilde\lambda)}_{n,k} =\Xi_*\,\mathbf{\Omega}_{n,k} \,.
\end{equation} 
Then let us fix 
\begin{equation}
 Y^A_\alpha= 
\begin{pmatrix} 
- y_\alpha^a\\
1_{(n-k)\times(n- k)}
\end{pmatrix} ,
\qquad \widetilde Y_{\dot \alpha}^{\dot A} = 
\begin{pmatrix} 
-\widetilde y^{\dot a}_{\dot \alpha}\\
1_{k\times k}
\end{pmatrix} \,,
\end{equation}
and write an explicit form of the orthogonal complements
\begin{equation}
( Y^\perp)^{ a}_{ A} = 
\left(\begin{tabular} {c|c}
$1_{2\times2}$&
$ y^{ a}_{ \alpha}$
\end{tabular} \right) ,
\qquad (\widetilde Y^\perp)^{\dot a}_{\dot A} = 
\left(\begin{tabular} {c|c}
$1_{2\times 2}$&
$\widetilde y^{\dot a}_{\dot \alpha}$
\end{tabular} \right)  \,.
\end{equation}
Moreover, we can decompose the matrices $\Lambda$ and $\widetilde\Lambda$ accordingly
\begin{equation}\label{big.lambda}
\Lambda_i^A = 
\begin{pmatrix} 
\lambda_i^{a *} \\
\Delta^\alpha_i
\end{pmatrix} ,
\qquad \widetilde\Lambda_i^{\dot A} = 
\begin{pmatrix} 
\widetilde\lambda_i^{\dot a *} \\
\widetilde\Delta^{\dot\alpha}_i
\end{pmatrix}  \,.
\end{equation}
Then \eqref{small.lambda} can be directly written as
\begin{equation}\label{subspace.lambda}
 \lambda_i^a = \lambda^{*a}_i+y_{\alpha}^a \Delta_i^\alpha \,,  \qquad\qquad
      \widetilde\lambda_i^{\dot a} = \widetilde\lambda^{*\dot{a}}_i+\widetilde y_{\dot{\alpha}}^{\dot{a}} \widetilde\Delta_i^{\dot\alpha} \,. 
\end{equation}
This discussion leads us to an alternative definition of the momentum amplituhedron $
\mathcal{M}^{(\lambda,\widetilde\lambda)}_{n,k}$, without any reference to auxiliary spaces. Let us define 
\begin{equation}\label{subspace.definition.mom}
\mathcal{V}_{n,k} 
= 
 \{(\lambda_i^a,\widetilde\lambda_i^{\dot a}):
     \lambda_i^a = \lambda^{*a}_i+y_{\alpha}^a \,\Delta_i^\alpha ,       \widetilde\lambda_i^{\dot a} = \widetilde\lambda^{*\dot{a}}_i+\widetilde y_{\dot{\alpha}}^{\dot{a}} \,\widetilde\Delta_i^{\dot\alpha},\lambda_i^a \widetilde\lambda_i^{\dot a}=0 
\}\,,
\end{equation}
where $(\lambda^*,\widetilde\lambda^*)$ are two fixed two-planes in $n$ dimensions, $\widetilde\Delta$ is a fixed $k$-plane and $\Delta$ is an $(n-k)$-dimensional fixed plane in $n$ dimensions. Moreover, we assume that when we assemble these subspaces as in \eqref{big.lambda},  $\widetilde\Lambda$ is a positive matrix and $\Lambda$ is a twisted positive matrix. Notice that $\mathcal{V}_{n,k}$ is a co-dimension-four subspace of an affine space of dimension $2n$.
We also define a winding space $\mathcal{W}_{n,k}$
\begin{align}
\mathcal{W}_{n,k}=&\{(\lambda_i^a,\widetilde\lambda_i^{\dot a}):\langle ii+1\rangle>0, [ii+1]>0\,, \nonumber \\&
\mbox{the sequence } \{\langle 12\rangle,\langle 13\rangle,\ldots,\langle 1n\rangle\} \mbox{ has } k-2 \mbox{ sign flips}\,, \nonumber \\
&\mbox{the sequence } \{[ 12],[ 13],\ldots,[ 1n]\} \mbox{ has } k \mbox{ sign flips}\} \,.
\end{align}
 Then the momentum amplituhedron $\mathcal{M}_{n,k}^{(\lambda,\widetilde\lambda)}$ in the spinor helicity space is the intersection:
$$
\mathcal{M}_{n,k}^{(\lambda,\widetilde\lambda)}=\mathcal{V}_{n,k}\cap \mathcal{W}_{n,k}\,.
$$

%%%%%%%%%%%%%%%%%%%%%%%%%%%%%%%%%%%%%%%%%%%%%%%%%%%%%%%%%%%%%

\subsubsection{Boundaries, Amplitude Singularities and Volume Form}

Before finding the volume form, \emph{i.e.}~the differential form with logarithmic singularities on all boundaries of $\mathcal{M}_{n,k}^{(m)}$, let us classify the boundaries. 
The boundaries of the momentum amplituhedron for $m=4$ have been recently studied in \cite{Ferro:2020lgp} using the  {\tt amplituhedronBoundaries} Mathematica\texttrademark~package   \cite{Lukowski:2020bya} and identified with relevant singularities of scattering amplitudes.
In particular, the facets of the momentum amplituhedron $\mathcal{M}_{n,k}^{(4)}$ 
belong to one of the following classes:
\begin{equation}
\label{boundaries}
\langle Y i i+1\rangle=0\,,\qquad [\widetilde Y ii+1]=0\,,\qquad S_{i,i+1,\ldots,j}=0\,.
\end{equation}
The first two classes can be related to all possible collinear limits of the amplitude.
The latter boundaries are written in terms of
\begin{equation}
S_{i,i+1,\ldots j}=\sum\limits_{a<b=i}^j \langle Y a b\rangle [\widetilde Y ab]\,,
\end{equation}
which are equivalent to the uplift of planar Mandelstam invariants to the momentum amplituhedron space.  These correspond to all possible non-trivial factorisations of the amplitude. The complete boundary stratification was found in \cite{Ferro:2020lgp} and each boundary element can be obtained by intersections of multiple facets, which translates into a combination of collinear limits and factorisations of amplitudes.

For $m=2$, the momentum amplituhedron $\mathcal{M}_{n,k}^{(2)}$ has the same boundary stratification as the hypersimplex $\Delta_{k+1,n}$, see \emph{e.g.}~\cite{Lukowski:2020bya}. In particular, the only facets of the momentum amplituhedron are of the form
\begin{equation}
\langle Yi\rangle=0\,,\qquad [\tilde Y i]=0 \,.
\end{equation}
Moreover, the facets of the former type are combinatorially equivalent to $\mathcal{M}_{n-1,k}^{(2)}$ and the ones of the latter type are equivalent to $\mathcal{M}_{n-1,k-1}^{(2)}$. This allows one to find the complete stratification of the momentum amplituhedron  $\mathcal{M}_{n,k}^{(2)}$  recursively.

The differential form $\mathbf{\Omega}^{(m)}_{n,k}$ with logarithmic singularities on all boundaries of the momentum amplituhedron can be found by 
 triangulating the space $\mathcal{M}_{n,k}^{(m)}$, with each triangle being an image through the map $\Phi_{(\Lambda,\widetilde\Lambda)}$ of a $\tfrac{m}{2}\cdot (n-\frac{m}{2})$-dimensional cell of the positive Grassmannian $G_{+}(k,n)$. To this extent,  the proper combination of cells can be found using the {\tt positroid} Mathematica\texttrademark~package \cite{Bourjaily:2012gy}.
 The logarithmic differential form on $\mathcal{M}_{n,k}^{(m)}$ is the sum over such cells of push-forwards of the canonical differential form for each cell. 
 The explicit answer is a sum of rational functions where the denominators can contain spurious singularities, corresponding to spurious boundaries in a given triangulation. These singularities disappear in the complete sum and the only divergences of $\mathbf{\Omega}^{(m)}_{n,k}$ correspond to the external boundaries.

%%%%%%%%%%%%%%%%%%%%%%%%%%%%%%%%%%%%%%%%%%%%%%%

\subsubsection{Integral representation}

One can also introduce a representation of the volume function $\Omega^{(m)}_{n,k}$ as an integral over a matrix space
 \begin{equation}
\delta^{\tfrac{m^2}{4}}(P) \,\Omega_{n,k} = \int 
\frac{\mbox{d}^{(n-k)\cdot (n-k)}{g}}{(\mbox{det}{g})^{n-k}}  \, \int_\gamma \omega_{n,k}  \prod_{\alpha=1}^{n-k} \delta^{(n-k+\tfrac{m}{2})}(Y-g\cdot c^\perp\cdot\Lambda)  \prod_{\dot\alpha=1}^{k} \delta^{(k+\tfrac{m}{2})}(\widetilde Y-\,c\cdot\widetilde\Lambda)\,,
\end{equation}
where we additionally need to integrate over the matrix $g$ corresponding to a $GL(n-k)$-transformation encoding the ambiguity of defining an orthogonal complement. The integration measure $\omega_{n,k}$ is the canonical measure on the space of $k\cdot n$ matrices $C$:
\begin{equation}
\omega_{n,k} = \frac{d^{k\cdot n}c_{\dot\alpha i}}{(12\ldots k)(23\ldots k+1)\ldots (n1\ldots k-1)}\,,
 \end{equation}
 where the brackets in the denominator are minors of the matrix $C$
\begin{equation}\label{c.minors}
(i_1 i_2\ldots i_k)=\epsilon_{\dot\alpha_1\dot\alpha_2\ldots\dot\alpha_k}\,c_{\dot\alpha_1i_1}c_{\dot\alpha_2i_2}\ldots c_{\dot\alpha_ki_k}\,.
\end{equation} 
The contour $\gamma$ can be found from \emph{e.g.}~BCFW recursion relations and it encircles a particular combination of poles of the integrand.

%%%%%%%%%%%%%%%%%%%%%%%%%%%%%%%%%%%%%%%%%%%%%%%%%%%%%%%

\subsubsection{Amplitudes from Momentum Amplituhedron}

Finally, we want to describe how to extract the amplitude $\mathcal{A}^{\text{\tiny tree}}_{n,k}$ from the volume form $\mathbf{\Omega}_{n,k}\equiv\mathbf{\Omega}_{n,k}^{(4)}$.  The momentum amplituhedron $\mathcal{M}_{n,k}$ is $(2n-4)$-dimensional and therefore the degree of $\mathbf{\Omega}_{n,k}$ is $(2n-4)$. 
Since the momentum amplituhedron is a subset of the $2n$-dimensional space $ G(n-k,n-k+2)\times G(k,k+2)$, then there are various ways one can write $\mathbf{\Omega}_{n,k}$ depending on the parametrisation of this subset. These different representations are related to each other by momentum conservation. In order to make the expression for the volume form independent of this choice, we use the fact that $1=\delta^4(P) d^4P$ and  define the volume function $\Omega_{n,k}$ in the following way:
\begin{equation}
\label{volumefc}
\mathbf{\Omega}_{n,k}  \wedge d^4P \, \delta^4(P) =\prod_{\alpha=1}^{n-k} \langle Y_1\ldots Y_{n-k} d^2 Y_\alpha \rangle \prod_{\dot{\alpha}=1}^{k} [ \widetilde Y_1\ldots \widetilde Y_{k} d^2 \widetilde Y_{\dot{\alpha}}]   \, \delta^4(P)\,\Omega_{n,k} \,.
\end{equation}
Indeed, the form $\mathbf{\Omega}_{n,k}  \wedge d^4P$ is top-dimensional and therefore can be written in terms of the measure on $ G(n-k,n-k+2)\times G(k,k+2) $ multiplied by a function.
Then, the procedure to extract the amplitude from the volume form $\mathbf{\Omega}_{n,k}$ is similar to the ordinary amplituhedron, \emph{i.e.}~we localise  $Y$ and $\widetilde Y$ on reference subspaces\footnote{This choice of $Y^*, \widetilde Y^*$ is compatible with the embedding of  $\lambda,\widetilde \lambda$ in  $\Lambda, \widetilde\Lambda$ as in \eqref{bosonizedY}, \eqref{bosonizedYt}.}
 \begin{equation}
 Y^*=\left(\begin{matrix}
\mathbb{0}_{2\times (n-k)}\\
\hline
\mathbb{1}_{(n-k)\times (n-k)}
\end{matrix}\right),\qquad\qquad 
\widetilde Y^*=\left(\begin{matrix}
\mathbb{0}_{2\times k}\\
\hline
\mathbb{1}_{k\times k}
\end{matrix}\right) \,.
\end{equation}  
We also introduce $2(n-k)$ auxiliary Grassmann-odd parameters $\phi_{a}^\alpha$, $\alpha=1,\ldots,n-k$ and $2k$ auxiliary Grassmann-odd parameters $\widetilde\phi_{\dot a}^{\dot\alpha}$, $\dot\alpha=1,\ldots,k$, defined as 
\begin{align}
\label{bosonizedY}
&\Lambda^A_i=\left(\begin{tabular}{c}$\lambda^a_i$\\$\phi_{a}^\alpha\cdot\eta^a_i$\end{tabular}\right),\qquad A=(a,\alpha)=1,\ldots,n-k+2\,,\\ 
\label{bosonizedYt} &\widetilde\Lambda^{\dot A}_i=\left(\begin{tabular}{c}$\widetilde\lambda^{\dot a}_i$\\$\widetilde\phi_{\dot a}^{\dot\alpha}\cdot\widetilde\eta^{\dot a}_i$\end{tabular}\right),\qquad \dot{A}=(\dot a,\dot\alpha)=1,\ldots,k+2\,.
\end{align} 
The amplitude can be found from
 \begin{equation}
 \label{extract}
 \mathcal{A}^{\text{\tiny tree}}_{n,k}=   \delta^4(p) \int d \phi^1_a \ldots  d\phi^{n-k}_a\int d\widetilde\phi^1_{\dot a}\ldots  d\widetilde\phi^k_{\dot a}\, \,\Omega_{n,k}(Y^*,\widetilde Y^*,\Lambda,\widetilde\Lambda) \,,
 \end{equation}
 where $\delta^4(p)$ comes from the localisation of $\delta^4(P)$ on $Y^*,\widetilde Y^*$.

 Alternatively, if we interpret the amplitude as a differential form on the spinor helicity space, we can extract the amplitude from the volume form $\mathbf{\Omega}^{(\lambda,\tilde\lambda)}_{n,k}$ in \eqref{omegaL} via the replacement
\begin{equation}
 \mathcal{A}^{\text{\tiny tree}}_{n,k}(\lambda,\tilde\lambda)=\mathbf{\Omega}_{n,k}^{(\lambda,\tilde\lambda)}\Big|_{d\lambda^a_i\to\eta^a_i,\, d\tilde\lambda^{\dot a}_i\to\tilde\eta^{\dot a}_i} \,.
\end{equation}

%%%%%%%%%%%%%%%%%%%%%%%%%%%%%%%%%%%%%%%%%%%%%%%
%%%%%%%%%%%%%%%%%%%%%%%%%%%%%%%%%%%%%%%%%%%%%%%
%%%%%%%%%%%%%ASSOCIAHEDRON%%%%%%%%%%%%%%%%%%%%%
%%%%%%%%%%%%%%%%%%%%%%%%%%%%%%%%%%%%%%%%%%%%%%%
%%%%%%%%%%%%%%%%%%%%%%%%%%%%%%%%%%%%%%%%%%%%%%%

\section{``Amplituhedra" For Bi-adjoint \texorpdfstring{$\phi^3$}{} Theory}
\label{sec:associahedron}

Positive geometries have been defined also for scattering amplitudes in other theories, beyond $\mathcal{N}=4$ sYM. In this section we will review the kinematic associahedron \cite{Arkani-Hamed:2017mur}, \emph{i.e.}~the ``amplituhedron" for $\phi^3$ theory, and  its close cousin, the worldsheet associahedron, which appears for open strings. We also discuss how the two are related by the ``scattering equations".

%%%%%%%%%%%%%%%%%%%%%%%%%%%%%%%%%%%%%%%%%%%%%%%
%%%%%%%%%%%%%%%%%%%%%%%%%%%%%%%%%%%%%%%%%%%%%%%
%%%%%%%%%%%%%%%%%%%%%%%%%%%%%%%%%%%%%%%%%%%%%%%

\subsection{Scattering amplitudes in \texorpdfstring{$\phi^3$}{} theory}

We start by discussing the scattering amplitudes in the bi-adjoint massless $\phi^3$ theory in $D$-dimensions, \emph{i.e.}~a theory of scalars in the adjoint representation of the product of two different color groups. The bi-adjoint color structure allows us to decompose an $n$-point amplitude into double-partial amplitudes $m_n(\alpha|\beta)$ labelled by two color orderings $\alpha$ and $\beta$, both given by a permutation of $n$ elements. 
From the point of view of geometry, most of the work has been done in the case when $\alpha=\beta$. Moreover, using cyclic symmetry of amplitudes we can subsequently focus on $m_n=m_n((12\ldots n)|(12\ldots n))$, where $(12\ldots n)$ indicates the standard ordering of $n$ elements. This introduces a particular fixed ordering between particles and restricts the class of diagrams one needs to consider to planar diagrams with respect to this ordering.

The double-partial amplitudes  are naturally written using Mandelstam variables
$$
s_{i_1\ldots i_r}=(p_{i_1}+\ldots p_{i_r})^2 \,,
$$
with  massless momenta $p_i^2=0$. Importantly, the Mandelstam variables are not linearly independent since the momenta $p_i$ satisfy the momentum conservation condition.
At tree level,  the amplitudes $m_n^{(0)}(\alpha|\alpha$) can be found by summing over all Feynman diagrams, which are color-ordered trivalent planar graphs, each contributing the product of its propagators\footnote{When the two permutations are different, the answer is a subset of the terms appearing in $m_n^{(0)}(\alpha|\alpha$).
}. They are therefore rational functions of Mandelstam variables. The positive geometry which describes them is the kinematic associahedron. At loop level, they become transcendental functions obtained from Feynman integrals. However, as in the previous section, there exists a positive geometry encoding the  integrands of Feynman integrals, at least at one loop: the halohedron.

%%%%%%%%%%%%%%%%%%%%%%%%%%%%%%%%%%%%%%%%%%%%%%%
%%%%%%%%%%%%%%%%%%%%%%%%%%%%%%%%%%%%%%%%%%%%%%%
%%%%%%%%%%%%%%%%%%%%%%%%%%%%%%%%%%%%%%%%%%%%%%%

\subsection{Kinematic Associahedron}
\label{sec:kinematic_associahedron}

As for $\mathcal{N}=4$ sYM, scattering amplitudes in the scalar bi-adjoint $\phi^3$ theory can be written as differential forms on the kinematic space. This suggests that one should look for a positive geometry directly in the kinematic space, without referring to any auxiliary construction, as an intersection of some positive region with an affine subspace. Such construction was proposed in \cite{Arkani-Hamed:2017mur} and the positive geometry obtained in this way 
is a projective version of the associahedron. 
The associahedron, also called  Stasheff polytope, is a well-known convex polytope of dimension $n-3$ which captures the combinatorics of subdivisions of an $n$-gon: each codimension $d$ boundary of the associahedron corresponds to a partial triangulation with $d$ diagonals inside an $n$-gon, and its interior corresponds to the trivial subdivision with no diagonals. The associahedron has a Catalan number $C_{n-2}$ of vertices and they correspond to the full triangulations of an $n$-gon. Alternatively, the vertices can be labelled by planar cubic tree graphs, dual to the triangulations. The Arkani-Hamed-Bai-He-Yan (ABHY) construction in \cite{Arkani-Hamed:2017mur} gives a particular realisation of the associahedron, directly in the kinematic space of Mandelstam invariants.

The associahedron naturally lives in the kinematic space $\mathcal{K}_n$ for $n$ massless particles in the bi-adjoint $\phi^3$ theory. This space is linearly spanned by
 the Mandelstam variables $s_{ij}$, which satisfy $n$ conditions of the form $\sum_{i\neq j} s_{ij} = 0$. Therefore, its dimension is $\mathrm{dim} \,\mathcal{K}_n = \frac{n (n-3)}{2}$.
There exists a natural choice for a basis of this space: given the standard ordering $(12\ldots n)$, one can define $\frac{n (n-3)}{2}$ planar variables 
\begin{equation}
X_{i,j} := s_{i,i+1,\ldots,j-1}\,,
\end{equation}
which are Mandelstam variables formed of momenta of consecutive particles, and which can be visualised as the diagonals between vertices $i$ and $j$ of a convex $n$-gon.

To define the kinematic associahedron we need two ingredients: a positive region and an affine space.
The positive region  $\Delta_n$ is defined by the requirement that all planar variables $X_{i,j}$ are positive 
\begin{equation}
X_{i,j}\geq 0\,,  \quad  \, 1\leq i< j\leq n \,.
\end{equation}
This defines a top-dimensional cone inside $\mathcal{K}_n$. The affine subspace is the $(n-3)$-dimensional subspace $H_n \subset \mathcal{K}_n$ defined by requiring that 
\begin{equation}
c_{i,j}=-s_{ij} =X_{i,j} + X_{i+1,j+1} - X_{i,j+1} - X_{i+1,j} \,,
\end{equation}
are positive constants for all non-adjacent $1 \leq i<j < n$. Notice that one does not restrict the variables $c_{i,j}$ when $j=n$. Then the kinematic associahedron $\mathcal{A}_n$ is defined as the intersection of the positive region $\Delta_n$ with the subspace $H_n$:
\begin{equation}
\mathcal{A}_n := \Delta_n \cap H_n \,.
\end{equation}
This is an $(n-3)$-dimensional subset of $\mathcal{K}_n$ which can be naturally parametrised by \emph{e.g.}~$X_{i,n}$ with $i=2,\ldots,n-2$. One can easily show that its boundary structure is identical to the $(n-3)$-dimensional associahedron. 
For instance, for four- and five-particle scattering we have:
\begin{eqnarray}
 \mathcal{A}_4 &=& \{s=X_{1,3}>0, t=X_{2,4}>0\} \cap \{-u=-s_{13}= \mathrm{const} >0\}\,, \\
 \mathcal{A}_5 &=& \{s_{12}=X_{1,3}>0, \ldots ,s_{51}=X_{2,5} > 0\} \cap \{-s_{13}, -s_{14}, -s_{24} =  \mathrm{const} >0\} \,.
\end{eqnarray}
 
The amplitudes can be now extracted from the unique canonical differential form on $\mathcal{A}_n$. Since the associahedron is a simple polytope, \emph{i.e.}~a $d$-dimensional polytope each of whose vertices are adjacent to exactly $d$ facets, the canonical form can be written as a sum over its vertices $v$ of the expressions $\pm \bigwedge\limits_{a=1}^{d} d\mathrm{log}F_a$, where $F_a=0$ describe the facets adjacent to $v$. For the kinematic associahedron all facets are characterised by the vanishing of one of the planar variables and therefore we can write:
\begin{equation}\label{eq:associahedronform}
\Omega(\mathcal{A}_n)= \sum_{p=1}^{C_{n-2}} \mathrm{sign}(v_p) \bigwedge_{a=1}^{n-3}d\mathrm{log} X_{i_a,j_a} \,.
\end{equation}
The signs $\mathrm{sign}(v_p)$ can be fixed by direct calculation of the canonical form, or by demanding that $\Omega(\mathcal{A}_n)$ is projective on $\mathcal{K}_n$.
One can show that this canonical form 
 computes the tree-level scattering amplitude $m_n^{(0)}$ for the bi-adjoint $\phi^3$ theory:
\begin{equation}
\label{eq:amplAssoc}
\Omega(\mathcal{A}_n)=  m_n^{(0)}  d^{n-3}X\,.
\end{equation}
For instance, for $n=4,5$ we have:
\begin{eqnarray}
\Omega(\mathcal{A}_4) &=& \left(\frac{dX_{1,3}}{X_{1,3}} - \frac{dX_{2,4}}{X_{2,4}}\right) = \left(\frac{1}{s} +\frac{1}{t} \right) ds   \,,   \\
\Omega(\mathcal{A}_5)  &=&  \left(\frac{1}{X_{1,3}X_{1,4}} +\ldots+ \frac{1}{X_{2,5}X_{3,5}}\right) dX_{2,5} \wedge dX_{3,5} \,.
\end{eqnarray}
This reproduces the results from Feynman calculations, where each term in the expansion comes from a planar trivalent graph.

We have already noticed that the canonical form of the kinematic associahedron can be found using the fact that it is a simple polytope: 
this leads to the representation \eqref{eq:associahedronform}. Other representations of canonical forms are also possible to find. For example,
a new recursion relation using a one-parameter deformation of kinematic variables $X_{ij} \rightarrow z X_{ij}$ has been provided  in \cite{He:2018svj}. By solving this recursion relation, one finds the bi-adjoint $\phi^3$ amplitudes in ``BCFW representation".

These results can also be generalised beyond the standard ordering, to partial amplitudes $m_n^{(0)}(\alpha | \beta)$. From the point of view of Feynman diagrams, only diagrams compatible with both orderings will contribute to the answer. From the geometric point of view, it will push some of the facets of the kinematic associahedron to infinity to obtain 
a different non-compact polyhedron for the various pairs of orderings \cite{Herderschee:2019wtl}. However, the canonical form of these polyhedra can be computed using a prescription equivalent to  \eqref{eq:amplAssoc} and encodes the partial amplitudes $m_n^{(0)}(\alpha | \beta)$. 

The ABHY construction can also be generalised to all polytopes associated with finite-type cluster algebras \cite{Bazier-Matte:2018rat}, where the usual associahedron corresponds to cluster algebras of type $A_n$. In particular, the canonical form of the $D_n$ associahedron gives the integrand for one-loop bi-adjoint $\phi^3$-amplitudes,
while the types $B_n$ or $C_n$ are related to one-loop diagrams with tadpole emissions. For all these polytopes, this construction has a natural physical origin coming from the $(1+1)$-dimensional causal
structure in kinematic space \cite{Arkani-Hamed:2019vag}. In this approach, the generalised associahedra become solutions to wave equations with positive source and their properties follow from simple properties of causal diamonds in the space-time.

Finally, we remark that there is a duality between differential forms on the kinematic space $\mathcal{K}_n$ and color factors. More specifically, the differential forms satisfy Jacobi relations similar to the usual Jacobi relations for structure constants, see \cite{Arkani-Hamed:2017mur}. This allows one to exchange kinematic factors with color factors, pointing towards a possible geometric motivation for the double-copy construction \cite{Bern:2008qj}.

%%%%%%%%%%%%%%%%%%%%%%%%%%%%%%%%%%%%%%%%%%%%%%%
%%%%%%%%%%%%%%%%%%%%%%%%%%%%%%%%%%%%%%%%%%%%%%%
%%%%%%%%%%%%%%%%%%%%%%%%%%%%%%%%%%%%%%%%%%%%%%%

\subsection{The Halohedron}

The kinematic associahedron captures the tree-level color-ordered scattering processes for bi-adjoint  $\phi^3$ theory for the standard ordering. A geometric construction can be also extended to include the integrand of one-loop bi-adjoint $\phi^3$ amplitudes and the positive geometry encoding this integrand is the halohedron   \cite{Salvatori:2018fjp,Salvatori:2018aha}.
The halohedron $\mathcal{H}_n$ is the convex polytope associated with the moduli space of  an annulus with marked points on one boundary \cite{devadoss2010deformations}\footnote{Note that this annulus is not associated to a cluster algebra, but the halohedron has a combinatorial structure very similar to $\overline{D}_n$, \emph{i.e.}~a polytope obtained by cutting the $D_n$ associahedron in half \cite{Arkani-Hamed:2019vag}.}. This generalises the moduli space of a disc which is associated with the tree-level construction. For one-loop, the marked points represent the external particles, while non-intersecting arcs, which generalise the $n$-gon diagonals we discussed for tree level, correspond to propagators of 1-loop planar diagrams. 
Then, the vertices of the halohedron are labelled by the planar one-loop Feynman diagrams, while the facets correspond to cuts of the one-loop integrand.

The halohedron is defined in an $n$-dimensional space $X$ with coordinates $(X_1,\ldots,X_n)$.  One can think of the space $X$ as the abstract kinematic space of all planar variables where the momentum conservation is not enforced. One starts by defining a set of linear functions $X_I$ which are in one-to-one correspondence with propagators of one-loop planar diagrams. Then the halohedron is defined as the region where all these variables are positive. This can be done by iterated truncations of an $n$-dimensional cube, as summarised in \cite{Salvatori:2018aha}.
As for the associahedron, the halohedron is a simple polytope and its logarithmic differential form can be found as
\begin{equation}
\Omega(\mathcal{H}_n) = \sum_{g} \text{sign}(g) \bigwedge_{I\in g}\frac{dX_I}{X_I} = d^nX \sum_{g} \prod_{I\in g} \frac{1}{X_I} \,,
\end{equation}
where the sum runs over all one-loop planar diagrams, including tadpoles and bubbles, and $I$ runs over all the propagators of a diagram $g$. 
The one-loop integrand for the bi-adjoint theory is obtained by killing the tadpole and bubble contributions by sending the corresponding variables $X_I$ to infinity, and going back to the physical kinematic space by substituting $X_I$ with the physical propagator $s_I$. In this way
momentum conservation is restored and  the logarithmic differential form $\Omega(\mathcal{H}_n)$ computes the integrand $m_n^{(1)}$ of the one-loop amplitude 
\begin{equation}
\Omega(\mathcal{H}_n)=m_n^{(1)} d^nX \,.
\end{equation}

%%%%%%%%%%%%%%%%%%%%%%%%%%%%%%%%%%%%%%%%%%%%%%%
%%%%%%%%%%%%%%%%%%%%%%%%%%%%%%%%%%%%%%%%%%%%%%%
%%%%%%%%%%%%%%%%%%%%%%%%%%%%%%%%%%%%%%%%%%%%%%%

\subsection{Worldsheet Associahedron and Scattering Equations}
\label{sec:worldsheetAssoc}

The associahedron plays a fundamental role also for open strings.
Indeed, the moduli space for the open-string worldsheet provides a different realisation of the associahedron. 
The open string moduli space is given by the real part $\mathcal{M}_{0,n}(\mathbb{R})$ of the moduli space of genus zero $\mathcal{M}_{0,n}$, which is the space of configurations of $n$ punctures on the Riemann sphere modulo $SL(2,\mathbb{C})$. It is equivalent to the moduli space of $n$ ordered points $\sigma_i$ on the boundary of a disc.
We also define the positive moduli space as the region associated with the standard ordering 
\begin{equation}
\mathcal{M}_{0,n}^+ := \{\sigma_1<\ldots<\sigma_n\}/SL(2,\mathbb{R}) \,.
\end{equation}
The Deligne-Mumford compactification of  $\mathcal{M}_{0,n}^+ $ \cite{Deligne69theirreducibility}, \emph{i.e.}~the blow-up of the open-string worldsheet  which makes manifest all the boundaries, has the same boundary structure as the associahedron and it is called the worldsheet associahedron. We will indicate the compactified space as $\overline{\mathcal{M}}_{0,n}^+ $.
Since $\overline{\mathcal{M}}_{0,n}^+ $ has the same boundary structure as the kinematic associahedron, its canonical form should be similar to \eqref{eq:associahedronform}.
Indeed, it can be shown that the canonical form on $\overline{\mathcal{M}}_{0,n}^+$ is 
\begin{equation}
\Omega_n(\overline{\mathcal{M}}_{0,n}^+ ) = \sum_{\mathrm{planar} \,g} \mathrm{sign}(g) \bigwedge_{a=1}^{n-3}d\mathrm{log} \left(\sigma_{i_a}-\sigma_{j_a-1}\right) \,,
\end{equation}
where the sum runs over all trivalent planar graphs, and for every $g$ the $(i_a, j_a)$ for $a = 1,\ldots , n-3$
are the diagonals of the corresponding triangulation. This can be further recast as a ``worldsheet Parke-Taylor" form 
\begin{equation}
\label{eq:omegaWS}
\Omega_n(\overline{\mathcal{M}}_{0,n}^+ )  = \frac{1}{\mathrm{vol}(SL(2))} \prod_{a=1}^n \frac{d\sigma_a}{\sigma_a-\sigma_{a+1}}\,.
\end{equation}

Finally, let us discuss the relation between the kinematic and worldsheet associahedra. The scattering equations relate points in the moduli space $\mathcal{M}_{0,n}$ to points in kinematic space $\mathcal{K}_n$ in the following way:
\begin{equation}
\sum_{j=1,j\neq i}^n \frac{s_{i,j}}{\sigma_i-\sigma_j}=0\,, \text{ for } i=\,\ldots,n\,.
\end{equation}
 As it is natural to expect, they also relate the two associahedra: on the subspace $H_n$ the scattering equations act as a diffeomorphism from the worldsheet associahedron $\overline{\mathcal{M}}_{0,n}^+ $ to the  kinematic associahedron $\mathcal{A}_n$. 
A diffeomorphism between two positive geometries implies the pushforward between the canonical forms \cite{Arkani-Hamed:2017tmz}, see Section \ref{sec:findforms}. Therefore, the scattering equation map pushes the canonical form of the worldsheet associahedron to that of the kinematic associahedron by summing over the  $(n-3)!$ solutions of the scattering equations:
\begin{equation}
\Omega_n(\overline{\mathcal{M}}_{0,n}^+ ) \xrightarrow[\text{scatt.eqs}]{\text{pushforward}} \Omega(\mathcal{A}_n) \,.
\end{equation}
Then, using \eqref{eq:amplAssoc} and \eqref{eq:omegaWS}, this implies  that the tree-level amplitude $m_n^{(0)}$ can be obtained by pushforward of the Parke-Taylor form via the scattering equations. 
This also provides a novel, geometric derivation of the CHY formula \cite{Cachazo:2013hca} for bi-adjoint scalars.

%%%%%%%%%%%%%%%%%%%%%%%%%%%%%%%%%%%%%%%%%%%%%%%
%%%%%%%%%%%%%%%%%%%%%%%%%%%%%%%%%%%%%%%%%%%%%%%
%%%%%%%%%%%%%%%OTHER GEOMETRIES%%%%%%%%%%%%%%%%
%%%%%%%%%%%%%%%%%%%%%%%%%%%%%%%%%%%%%%%%%%%%%%%
%%%%%%%%%%%%%%%%%%%%%%%%%%%%%%%%%%%%%%%%%%%%%%%

\section{Other Positive Geometries}
\label{sec:other}

In this section we will briefly review other positive geometries which have been found in recent years for other observables in physics. In particular, we describe the notions of the {cosmological polytope}
	\cite{Arkani-Hamed:2017fdk,Arkani-Hamed:2018bjr,Benincasa:2018ssx} and  the geometrical structure underlying the conformal bootstrap program \cite{Arkani-Hamed:2018ign}.

%%%%%%%%%%%%%%%%%%%%%%%%%%%%%%%%%%%%%%%%%%%%%%%
%%%%%%%%%%%%%%%%%%%%%%%%%%%%%%%%%%%%%%%%%%%%%%%
%%%%%%%%%%%%%%%%%%%%%%%%%%%%%%%%%%%%%%%%%%%%%%%

\subsection{Cosmological Polytopes}

Positive geometries have made their appearance also in cosmology: the cosmological polytope gives a connection between the wavefunction of the universe and polyhedral geometry, analogous to the one seen for scattering amplitudes \cite{Arkani-Hamed:2017fdk,Arkani-Hamed:2018bjr,Benincasa:2018ssx,Benincasa:2019vqr}. As with scattering amplitudes, the canonical form with logarithmic singularities on all the boundaries of this polytope computes the cosmological wavefunction.

Lorentz invariance is broken at cosmological scales and makes it impossible to have well-defined quantum mechanical observables.
However, under the assumption that the universe becomes infinitely large and flat at sufficiently late times, the late-time spatial correlation functions or, equivalently, the wavefunction of the universe generating them, are well-defined observables. 
Focusing on scalar fields, the momentum space correlators are
\begin{equation}
\langle \prod_{j=1}^n \phi(\overrightarrow{p_j}) \rangle = \int D\phi  \, \prod_{j=1}^n \phi(\overrightarrow{p_j})  |\Psi[\phi]|^2\,,
\end{equation}
where $\Psi[\phi]$ is the wavefunction of the universe and it has a representation in terms of path integrals. 
Let us consider a class of toy models of massless scalar fields in $(d+1)$ dimensions  with time-dependent polynomial interactions
\begin{equation}
\label{S_cosmo}
S[\phi] = \int_{-\infty}^0 d\eta \int d^dx \left[ \frac{1}{2} (\partial\phi)^2 - \sum_{k\geq 3} \frac{\lambda_k(\eta)}{k!} \phi^k\right] \,,
\end{equation}
where $\lambda_k(\eta)$ is the time-dependent coupling constant. 
The class of theories \eqref{S_cosmo} includes  as a special case  
conformally-coupled scalars with non-conformal polynomial interaction in Friedmann-Robertson-Walker (FRW) cosmologies.
The wavefunction can be computed perturbatively via Feynman diagrams. In these simple models, the result is a rational function of the sum $x_i = \sum_{k\in v_i} E_k$ of the energies of external states $E_k = |\vec{p}_k|$ at each vertex $v_i$ of the graph and on the internal energies $y_{ij}$ associated with the edges between the vertices $v_i$ and $v_j$. To a given Feynman graph $\mathcal{G}$ we can associate its  contribution to the perturbative wavefunction ${\psi}_{\mathcal{G}}(x_v,y_e)$
\begin{equation}
{\psi}_{\mathcal{G}}(x_v,y_e) = \int_{-\infty}^0 \prod_{v\in \mathcal{V}} d\eta_v e^{i x_v \eta_v} \prod_{e\in \mathcal{E}} G(\eta_v,\eta_{v'},y_e)\,,
\end{equation}
where $G(\eta_v,\eta_{v'},y_e)$ is the bulk-to-bulk propagator, while $ \mathcal{V}$ is the set of vertices and $ \mathcal{E}$ is the set of edges of the graph $ \mathcal{G}$. 
For these models it has been shown that the contribution of each Feynman diagram $\mathcal{G}$ to the perturbative wavefunction at all orders ${\psi}_{\mathcal{G}}(x_v,y_e)$ is related to the canonical form of a polytope, the cosmological polytope $\mathcal{P}$, which has an intrinsic definition without any reference to space-time. 
In particular, to any diagram ${\mathcal{G}}$ we can associate 
 vectors ${\bf{x}}_v$ with all the vertices and ${\bf{y}}_e$ with all the edges. 
These vectors give a basis for the projective space $\mathbb{P}^{n_e+n_v-1}$. 
To any graph we can associate a collection of intersecting triangles in the following way.  To each edge ${\bf{y}}_i$ with its two vertices  ${\bf{x}}_i$ and ${\bf{x}}'_i$  we associate  a triangle whose midpoints are identified by the vectors $({\bf{x}}_i, {\bf{x}}'_i, {\bf{y}}_i)$ -- the vertices of the triangle are therefore $\{{\bf{x}}_i + {\bf{x}}'_i - {\bf{y}}_i, {\bf{x}}_i - {\bf{x}}'_i + {\bf{y}}_i, -{\bf{x}}_i + {\bf{x}}'_i + {\bf{y}}_i\}$. The vertices  ${\bf{x}}_i, {\bf{x}}'_i$ of the graph  represent the two sides  on which the triangle can intersect other triangles. On the third edge of the triangle, with midpoint ${\bf{y}}_i$, no intersection is allowed.
The cosmological polytope $\mathcal{P}$ is the convex hull of the 3 $n_e$ vertices of $n_e$ intersected triangles.
Let us write any point in $\mathcal{P}$ as
\begin{equation}
\mathcal{Y} = \sum_{v} x_v {\bf{X}}_v + \sum_{e} y_e {\bf{Y}}_e \quad ({\bf{X}}_v,{\bf{Y}}_e,) \in \mathbb{R}^{n_v+n_e}\,,
\end{equation}
with $ ({\bf{X}}_v,{\bf{Y}}_e,)$ identifying the independent midpoints ${\bf{x}}$ and ${\bf{y}}$ of the triangles generating $\mathcal{P}$.
The coefficients $x_v$ and $y_e$ will label the vertices and the edges of the graph $\mathcal{G}$, and are not vertices of the cosmological polytope. 
Then, for a graph $\mathcal{G}$  one can associate a cosmological polytope $\mathcal{P}_{\mathcal{G}}$ with a logarithmic differential form associated to the wavefunction ${\psi}_{\mathcal{G}}(x_v,y_e)$:
\begin{equation}
\Omega(\mathcal{Y},\mathcal{P}) = \prod_{v\in \mathcal{V}} \prod_{e\in \mathcal{E}} dx_v dy_e {\psi}_{\mathcal{G}}(x_v,y_e)\,.
\end{equation}
The  boundaries of this geometry are lower-dimensional polytopes encoding the residues of the wavefunction poles. The triangulations are different representations of $\psi$.

Remarkably, the physics of the flat-space S-matrix is naturally contained in this object: a particular co-dimension one boundary related to the total energy pole $\sum_{i=1}^n \,E_i \,\rightarrow 0$, the so-called scattering facet, encodes the information of flat-space scattering amplitudes $A^{\mathrm{flat}} $. 
The scattering-facet structure encodes unitarity, in the way its boundaries factorise into products of lower-dimensional polytopes, and  Lorentz invariance, from the contour integral representation of its canonical form. 
Furthermore, for these toy models, it is possible to reconstruct the perturbative wavefunction from the knowledge of the flat-space amplitudes and the requirement of the absence of unphysical singularities \cite{Benincasa:2018ssx}.

Recently, this construction has been extended to a class of toy models of light massive scalars with time-dependent masses and polynomial couplings, which contains general massive scalars in FRW cosmologies  \cite{Benincasa:2019vqr}. The wavefunction of the universe is a degenerate limit of the canonical form of a particular generalisation of the cosmological polytopes described above. 

At the moment the cosmological polytopes describe each Feynman diagram separately while one would rather prefer a single geometry, providing compact expressions for the wavefunction. Nevertheless, these objects are a first step towards defining geometries analogous to amplituhedra and associahedra.

%%%%%%%%%%%%%%%%%%%%%%%%%%%%%%%%%%%%%%%%%%%%%%%
%%%%%%%%%%%%%%%%%%%%%%%%%%%%%%%%%%%%%%%%%%%%%%%
%%%%%%%%%%%%%%%%%%%%%%%%%%%%%%%%%%%%%%%%%%%%%%%

\subsection{Positive Geometry for Conformal Bootstrap}

Positive geometries are also arising in more general conformal field theories (CFTs), beyond $\mathcal{N}=4$ sYM. In particular, it is possible to translate the conformal bootstrap equation using geometric ideas we explored in previous sections,
 leading to new insights into the four-point functions in general CFTs \cite{Arkani-Hamed:2018ign}. In this geometric picture, unitarity demands that the partial-wave expansion coefficients of a four-point function lie inside a famous polytope called the cyclic polytope, and crossing symmetry restricts them to lie on a plane. Then the spectrum of CFTs can 
  be studied by investigating the rich geometric and combinatorial structures of the intersection of the plane with the polytope.

As an example of these ideas, we consider a unitary one-dimensional CFT and study a four-point function of identical, real conformal primary operators $\phi$ with scaling dimension $\Delta_{\phi}$. The $SL(2,\mathbb{R})$ covariance of the four-point function implies that it can be written as 
\begin{equation}
\langle \phi(x_1)\,\phi(x_2)\,\phi(x_3)\,\phi(x_4)\rangle = \frac{1}{|x_{12}|^{2\Delta_\phi} |x_{34}|^{2\Delta_\phi}} \,F(z) \,,
\end{equation}
where $F(z)$ is a function of the cross-ratio $z=\frac{x_{12}x_{34}}{x_{13}x_{24}}$. Taking the operator product expansion (OPE) of the operators $\phi(x_1)$ and $\phi(x_2)$, the function $F(z)$ can be written in terms of partial waves as
\begin{equation}
\label{eq:unitCFT}
F(z) = \sum_i p_i G_{\Delta_i}(z) \,, \quad p_i > 0 \,,
\end{equation}
where the coefficients $p_i$'s are positive due to unitarity. Here, the functions $G_{\Delta}(z)$ are the $SL(2,\mathbb{R})$ conformal blocks
\begin{equation}
G_{\Delta}(z) = z^{\Delta}\, {}_2F_1(\Delta,\Delta,2\Delta,z)  \,.
\end{equation}
Comparing \eqref{eq:unitCFT} with the expression found by computing the OPE of the operators $\phi(x_2)$ and $\phi(x_3)$, one finds the crossing equation:
\begin{equation}
\label{eq:crossCFT}
F(z) = \left( \frac{z}{1-z}\right)^{2\Delta_{\phi}} F(1-z)\,.
\end{equation}
The conformal bootstrap program aims to study the space of solutions of $\Delta_i$ and $p_i$ by finding a solution to the unitarity  and crossing equations, \eqref{eq:unitCFT} and \eqref{eq:crossCFT} respectively. This infinite-dimensional problem can be approached by discretizing the four-point function: instead of considering the complete function $F(z)$, one takes a truncation of its Taylor expansion around $z=\frac{1}{2}$ to the first $2N+1$ derivatives:
\begin{equation}
{\bf{F}}= \left(
\begin{array}{c}
F_\Delta^0 \\
F_\Delta^1 \\
\vdots \\
F_\Delta^{2N+1}
\end{array}
\right) \, \in \mathbb{P}^{2N+1}\,,
\end{equation}
with $F^I \equiv \frac{1}{I!}\partial_z^I F(z)|_{z=\frac{1}{2}}$, for $I=1,2,\ldots,2N+1$.
The same can be done for the conformal block $G_{\Delta}(z)$: its Taylor expansion around $z=\frac{1}{2}$ gives  a $(2N + 2)$-dimensional block vector ${\bf{G}}_{\Delta}$
\begin{equation}
{\bf{G}}_{\Delta} = \left(
\begin{array}{c}
G_\Delta^0 \\
G_\Delta^1 \\
\vdots \\
G_\Delta^{2N+1}
\end{array}
\right) \, \in \mathbb{P}^{2N+1}\,.
\end{equation}
The unitarity condition \eqref{eq:unitCFT} demands that the Taylor coefficients of the four-point function $F(z)$ expanded around $z=\frac{1}{2}$ have to lie 
 in the positive span of the block vectors ${\bf{G}}_{\Delta}$, \emph{i.e.}~inside a polytope spanned by the block vectors:
 \begin{equation}
 {\bf{F}} = \sum_\Delta p_\Delta {\bf{G}}_\Delta \,, \quad p_\Delta > 0 \,.
 \end{equation}
 This polytope is called the unitarity polytope ${\bf{U}}[\{\Delta_i\}]$. Since one can show that the determinant $\langle {\bf{G}}_{\Delta_1}{\bf{G}}_{\Delta_2} \ldots {\bf{G}}_{\Delta_n}\rangle$ for $\Delta_1<\Delta_2<\ldots<\Delta_n$ is positive\footnote{There is a  caveat that for sufficiently small $\Delta$'s and large $d$, the minors can be negative. This is irrelevant from a practical point of view, due to the fact that  the negative minors are always extremely small.}, one finds that ${\bf{U}}[\{\Delta_i\}]$ is a cyclic polytope.
On the other hand, the crossing equation \eqref{eq:crossCFT} restricts the  Taylor coefficients of $F(z)$ to lie on an $N$-dimensional plane, called the crossing plane ${\bf{X}}[\Delta_{\phi}]$, which  is fixed by the dimension $\Delta_\phi$ of the scalar operator $\phi$.
A  four-point function is consistent with unitarity and crossing if the coefficients $p_\Delta$ lie in the region defined by the intersection of the fixed $N$-dimensional crossing plane ${\bf{X}}[\Delta_{\phi}]$ and the $(2N + 1)$-dimensional unitarity polytope ${\bf{U}}[\{\Delta_i\}]$ which varies with the spectrum. One immediate implication of this construction is that a consistent CFT must contain an infinite number of operators in its spectrum. 

The above construction implies that finding a solution to the conformal bootstrap equation corresponds geometrically to demanding that the intersection of ${\bf{U}}[\{\Delta_i\}] \cap {\bf{X}}[\Delta_{\phi}]$ is not empty. This provides bounds on the four-point function and allows the identification of the space of consistent CFTs  geometrically. Since the face structure of cyclic polytopes is completely understood, one can fully characterise the intersection combinatorially. 
This allows one to find new exact statements about the spectrum and four-point function in any CFT. For instance, one can show that when the spectrum is continuously varied the shape of the intersection changes, which may lead to various discrete jumps in the geometry, akin to ``phase transitions".

In  \cite{Arkani-Hamed:2018ign},  this geometry has been investigated in details for the cases when $N = 1, 2$. This allowed for rigorous study of how the space of consistent $\Delta$'s is carved out by the bootstrap at this resolution.  Going to higher $N$, which means keeping more terms in the Taylor expansion, the resolution on CFT data increases, providing a further refinement of the space of allowed operator dimensions. This leads to an efficient procedure increasing the resolution and allows one to build up the space of allowed $\Delta$'s recursively.

%%%%%%%%%%%%%%%%%%%%%%%%%%%%%%%%%%%%%%%%%%%%%%%
%%%%%%%%%%%%%%%%%%%%%%%%%%%%%%%%%%%%%%%%%%%%%%%
%%%%%%%%%%%%%%%%%RECENT ADVANCES%%%%%%%%%%%%%%%
%%%%%%%%%%%%%%%%%%%%%%%%%%%%%%%%%%%%%%%%%%%%%%%
%%%%%%%%%%%%%%%%%%%%%%%%%%%%%%%%%%%%%%%%%%%%%%%

\section{Recent Advances}
\label{sec:recent}
In this section, we wish to briefly review the most recent progress made related to positive geometries and their extensions. In particular, we will first discuss the deformation of logarithmic differential forms of polytopes which give rise to the stringy canonical forms.
Afterwards, we comment on various relations of positive geometries to tropical geometry.

In the previous sections we described positive geometries and rational forms which can be naturally assigned to them. Many answers in high-energy physics are however given by transcendental functions rather than rational ones, for example when studying loop scattering amplitudes, or string theory amplitudes. To accommodate for them one needs to expand the geometric description and allow algebraic structures beyond logarithmic differential forms. One possible extension was given in \cite{Benincasa:2020uph} where forms with higher-order poles have been considered. Another direction was presented in   
\cite{Arkani-Hamed:2019mrd}, where an $\alpha'$-extension of canonical forms for polytopes has been introduced, with $\alpha'$ reminiscent of the string theory parameter.  
These so-called ``stringy integrals" share various properties with string amplitudes and  are defined as integrals of logarithmic forms regulated by polynomials with exponents.
They have the natural property that when $\alpha'\rightarrow 0$  they reduce to the usual canonical form of a polytope given by the Minkowski sum of the Newton polytopes of the regulating polynomials. From the string theory point of view this would be called the field-theory limit.  Moreover, when one considers the $\alpha'\rightarrow \infty$ limit, the saddle-point equations for the stringy integrals  give the scattering equations. These provide a diffeomorphism from the integration domain to the polytope, and therefore a pushforward formula for its canonical form.
Finally, at finite $\alpha'$ the stringy integrals have simple poles corresponding to facets of the polytope and the residue evaluated at the pole is given by the stringy canonical form of the facet.  This provides a natural generalisation of the property in the definition of positive geometries.

The stringy integrals can be defined for any polytope and therefore provide extensions of the logarithmic differential forms for positive geometries in the second class of the classification described in section \ref{sec:physics}. 
One starts by considering the integral over  $\mathbb{R}_+^d = \{0<x_i<\infty\}$ of the canonical form of a simplex $\prod_{i=1}^d \frac{dx_i}{x_i}$. Such integral is divergent when $x_i\rightarrow 0$ and $x_i\rightarrow \infty$ and to regulate these divergences one introduces the parameters $X_i>0$, $i=1,\ldots,d$ and $c>0$ and considers the following integral:
\begin{equation}
\label{eq:stringy}
\mathcal{I}_p({\bf{X}},c) := (\alpha')^d \int_0^{\infty} \prod_{i=1}^d \frac{dx_i}{x_i} x_i^{\alpha' X_i} p({\bf{x}})^{-\alpha' c} \,.
\end{equation}
Here ${\bf{X}}=(X_1,\ldots,X_d)$, ${\bf x}=(x_1,\ldots,x_d)$ and $p({\bf{x}})$ is a polynomial with positive coefficients:
\begin{equation}
p({\bf{x}}) := \sum_{\alpha} \, p_\alpha \, {\bf{x}}^{{\bf{n}}_i}\,,
\end{equation}
where $p_\alpha>0$ and ${\bf{x}}^{{\bf{n}}_i}:= x_1^{n_{i,1}} \ldots x_d^{n_{i,d}}$.
The integral $\mathcal{I}_p$ converges if and only if  the Newton polytope of the polynomial $p(\bf x)$, \emph{i.e.}~the convex hull of the exponent vectors ${\bf{n}}_i \in \mathbb{Z}^d$:
\begin{equation}
 N[p] =\left\{ \sum_{\alpha} \lambda_\alpha {\bf{n}}_\alpha :\, \lambda_\alpha \geq 0 \,,  \sum_\alpha \lambda_\alpha  =1 \right\}\,,
\end{equation}
 is $d$-dimensional and ${\bf{X}}$ is inside the polytope $c N[p]$. 
 
Importantly, the $\alpha'\rightarrow 0$ limit of stringy integrals gives the canonical form of the (rescaled) Newton polytope:
\begin{equation}
\underset{{\alpha'\rightarrow 0}}{\mathrm{lim}}\mathcal{I}_p({\bf X},c)d^d{\bf X} = \mathbf{\Omega}(c N[p]) \,.
\end{equation}
On the other hand, if we consider the limit $\alpha'\rightarrow \infty$ then the saddle-point equations obtained from the integral \eqref{eq:stringy}: 
\begin{eqnarray}
\label{eq:SE}
 X_i = x_i \frac{c}{p({\bf{x}})} \frac{\partial p({\bf{x}})}{\partial x_i}
\end{eqnarray}
provide a diffeomorphism $\Phi$ from $\mathbb{R}_+^d $ to the interior of the polytope  $c N[p]$:
\begin{eqnarray}
\Phi &:& \mathbb{R}_+^d \rightarrow c N[p] \,.
\end{eqnarray}
We can use this diffeomorphism to perform a pushforward of the simplex canonical form  and obtain the canonical form of the Newton polytope  
\begin{equation}
\Phi_* \left(\prod_{i=1}^d \frac{dx_i}{x_i}\right) = \mathbf{\Omega}(cN[p]) \,.
\end{equation}
These provide two alternative methods to find canonical forms for polytopes which can be realised as Newton polytopes. This can be interpreted as the statement that, for any polytope, the low-energy limit of the stringy canonical form agrees with the pushforward using the scattering equations from the saddle points in the high-energy limit.

This construction can be generalised to the case with multiple subtraction-free Laurent polynomials $p_I(\bf x)$ and regulating parameters $c_I$. 
Such integrals converge when ${\bf{X}}$ is inside $\mathcal{P}:=\bigoplus_I c_I N(P_I)$ -- the Minkowski sum of Newton polytopes for each polynomial.
As before,  the leading order $\alpha'\rightarrow 0$ of these integrals is the canonical form on the Minkowski sum $\mathbf{\Omega}(\mathcal{P})$. Moreover, the saddle point equations as $\alpha'\to \infty$ provide a diffeomorphism from the simplex to $\mathcal{P}$. This allows for an alternative way to find  the canonical form on $\mathcal{P}$ using the pushforward of the simplex canonical form.

As an example, one can apply the stringy integrals to the ABHY associahedron $\mathcal{A}_{n}$ which we described in section \ref{sec:kinematic_associahedron}. In particular, the associahedron can naturally be represented as a Minkowski sum of simpler polytopes. This decomposition provides a particular choice of regulating polynomials. Using these polynomials, the stringy integral associated to the associahedron reproduces the usual open-string integral with the Koba-Nielsen factor as regulator:
\begin{equation}\label{eq:openstring}
\mathcal{I}_n^{\text{disk}}:= (\alpha')^{n-3} \int_{\overline{\mathcal{M}}^+_{0,n}} \Omega(\overline{\mathcal{M}}^+_{0,n}) \prod_{a<b} |z_a-z_b|^{\alpha' s_{ab}} \,.
\end{equation}
This provides a direct path from kinematic space to string amplitudes without any reference to the string worldsheet or space-time.
As before, the field-theory limit $\alpha'\rightarrow 0$ of $\mathcal{I}_n^{\text{disk}}$ is the canonical function of the associahedron $\Omega(\mathcal{A}_{n})$, which encodes the bi-adjoint $\phi^3$ tree-level $n$-particle amplitude. The latter can also be computed by performing the pushforward of $\Omega(\overline{\mathcal{M}}^+_{0,n})$ using the CHY scattering equations, which are the saddle-points of the Koba-Nielsen factor from the Gross-Mende limit $\alpha'\rightarrow \infty$.

Similar integrals can also be constructed starting from the generalised cluster associahedra \cite{Bazier-Matte:2018rat} to obtain general cluster string integrals. 
 These are reminiscent of the ordinary string theory scattering amplitudes, which correspond to cluster algebras of type $A_n$, with properties relevant for scattering of generalised particles and strings. Stringy canonical forms have also been studied for generalised permutohedra \cite{He:2020onr}. 
In both cases of generalised associahedra and permutohedra, the combinatorics of these polytopes can be explored using the idea of binary geometries \cite{Arkani-Hamed:2019plo}.
Moreover, the stringy canonical forms can further be extended beyond polytopes to Grassmannian string integrals, \emph{i.e.}~integrals over the  positive Grassmannian modulo torus action $G_+(k,n)/T$ \cite{Arkani-Hamed:2019rds,He:2020ray}.

As described above, the stringy canonical forms are convergent if the exponents satisfy particular positivity conditions. It is however possible to extend the notion of stringy forms to include all exponents by using the methods of tropical geometry. Tropical geometries, and in particular the tropical Grassmannians, have recently made multiple appearances in the context of scattering amplitudes. They are related to extensions of the bi-adjoint scalar theories described in section \ref{sec:associahedron}. In this context, the associahedron is related to a configuration space of $n$  points on the projective space $\mathbb{CP}^1$. It is captured by the positive tropical Grassmannian $\text{Trop}_+Gr(2,n)$ --  the space of phylogenetic trees which can be associated to Feynman diagrams for the $\phi^3$ theory. A more general class of theories proposed in \cite{Cachazo:2019ngv} describe configuration spaces of points on $\mathbb{CP}^{k-1}$ and are governed by its generalisation:  $\text{Trop}_+Gr(k,n)$ \cite{troppos}. Tropical Grassmannians also play a prominent role in the discussions on symbol alphabets for loop amplitudes in planar $\mathcal{N}=4$ sYM \cite{Arkani-Hamed:2019rds,Drummond:2019cxm,Drummond:2020kqg,Henke:2019hve} and their relation to cluster algebras. Finally, the positive tropical Grassmannian $\text{Trop}_+Gr(k+1,n)$ governs positroid dissections of the hypersimplex \cite{Lukowski:2020dpn,Early:2019eun,Arkani-Hamed:2020cig}, and therefore are related to a particular class of triangulations of the amplituhedron $\mathcal{A}_{n,k}^{(2)}$ through the T-duality map \cite{Lukowski:2020dpn}.  
%

%%%%%%%%%%%%%%%%%%%%%%%%%%%%%%%%%%%%%%%%%%%%%%%
%%%%%%%%%%%%%%%%%%%%%%%%%%%%%%%%%%%%%%%%%%%%%%%
%%%%%%%%%%%%%%%%%%CONCLUSIONS%%%%%%%%%%%%%%%%%%
%%%%%%%%%%%%%%%%%%%%%%%%%%%%%%%%%%%%%%%%%%%%%%%
%%%%%%%%%%%%%%%%%%%%%%%%%%%%%%%%%%%%%%%%%%%%%%%

\section{Conclusions and Open Problems}
In this review we summarised recent developments in geometric descriptions of observables in physics, with a special emphasis on positive geometries relevant for scattering amplitudes: amplituhedra. Positive geometries provide a completely new framework for computing and understanding physical quantities, and a plethora of -- some of which previously hidden --  properties can be extracted by studying the structure of these geometrical objects.

We are only at the beginning of our journey towards a complete understanding of positive geometries, and there is a large number of open questions which will keep both physicists and mathematicians busy in the years to come. In the following, we compile a (non-comprehensive) list of the most significant open problems and challenges.

%%%%%%%%%%%%%%%%%%%%%%%%%%%%%%%%%%%%%%%%%%%%%%%

\paragraph{Understanding known geometries.}
Despite the great progress which has been achieved in the understanding of known geometries, even for the oldest example, the amplituhedron, much of our knowledge comes from a case-by-case study and we lack general statements. 
Among the various interesting questions which are still open, the following are in our opinion the most pressing: 
\begin{itemize}[leftmargin=*]
\item[{$\star$}] Can we produce compact, closed expressions for the canonical forms of known geometries?
\item[{$\star$}] Can we classify all the triangulations of positive geometries to get access to all possible representations of a given observable?
\item[{$\star$}] Can we provide a combinatorial description of all boundaries of the geometries to understand  and classify all possible physical singularities of a given observable?
\end{itemize}
Some of these questions, as we described in the main text, have been already (partially) answered for some of the positive geometries we know, but for many others little is known in these respects.

%%%%%%%%%%%%%%%%%%%%%%%%%%%%%%%%%%%%%%%%%%%%%%%

\paragraph{Finding new geometries.}

The process of finding new positive geometries relevant for physics is still on its way.  
Some of these we expect to exist but we do not have a direct construction yet. In particular, the geometries which are sought-after at the moment are:
\begin{itemize}[leftmargin=*]
\item[{$\star$}] Loop momentum amplituhedron: a geometry encoding the integrand for $\mathcal{N}=4$ sYM scattering amplitudes directly in the spinor helicity space. One immediate problem with finding this geometry is the ambiguity in defining the loop momentum using spinor helicity variables.
\item[{$\star$}] Positive geometries for non-planar theories. The amplituhedron has an ordering built-in into its definition as it is formulated in momentum twistor space. One needs therefore to use spinor helicity (or twistor) variables to discuss an extension of geometries beyond the planar sector. This makes the momentum amplituhedron a good starting point for such extensions. 
\item[{$\star$}] Geometries for scattering amplitudes in more realistic theories, including Quantum Chromodynamics.
\item[{$\star$}] Loop associahedron: a geometry encoding all-loop integrands of biadjoint $\phi^3$ theory.
\item[{$\star$}] A single geometry underlying the wavefunction of universe: such geometry would allow us to find the wavefunction of the universe in a single step, without referring to many cosmological polytopes contributing to it. 
\end{itemize}

%%%%%%%%%%%%%%%%%%%%%%%%%%%%%%%%%%%%%%%%%%%%%%%

\paragraph{Beyond the integrand.}

Since positive geometries are naturally associated with rational functions, they provide us with integrands rather than integrals for scattering amplitudes at loop level. In section \ref{sec:recent} we explained how this has recently changed, with new methods available to associate transcendental functions to positive geometries. The main question is then whether we can extract the integrated amplitudes directly from the underlying geometry.

%%%%%%%%%%%%%%%%%%%%%%%%%%%%%%%%%%%%%%%%%%%%%%%

\paragraph{Mathematical precision.}

Many results available for positive geometries  are until now based on case-by-case studies and are often strongly rooted in physics intuition. This is not satisfactory from a mathematical point of view. Some of the basic questions for which we lack a rigorous mathematical proof include:
\begin{itemize}[leftmargin=*]
\item[{$\star$}] Are the amplituhedron and the momentum amplituhedron positive geometries?
\item[{$\star$}] Do amplituhedra admit triangulations and is the BCFW triangulation one of them?
\item[{$\star$}] Are the amplituhedron and the momentum amplituhedron homeomorphic to a ball?
\end{itemize}

\noindent
More generally, positive geometries provide a novel framework for quantum field theory where locality and unitarity are emergent concepts, and positivity replaces them as the main axiom. One of the main questions is then whether we can completely avoid introducing Lagrangians and gauge dependent methods and re-derive all known results, as well as not known ones, using only well-defined, geometric, not redundant methods without ever mentioning Feynman integrals.

%%%%%%%%%%%%%%%%%%%%%%%%%%%%%%%%%%%%%%%%%%%%%%%
%%%%%%%%%%%%%%%%%%%%%%%%%%%%%%%%%%%%%%%%%%%%%%%
%%%%%%%%%%%%%%%%%%%%%%%%%%%%%%%%%%%%%%%%%%%%%%%
%%%%%%%%%%%%%%%%%%%%%%%%%%%%%%%%%%%%%%%%%%%%%%%
%%%%%%%%%%%%%%%%%%%%%%%%%%%%%%%%%%%%%%%%%%%%%%%

\section{Acknowledgements}
We would like to thank D. Damgaard, C. Meneghelli, R. Moerman, A. Orta, M. Parisi, J. Plefka, M. Spradlin, M. Staudacher, A. Volovich,  L. Williams for collaboration on the topics presented in this review and N. Arkani-Hamed, P. Benincasa, {\"O}. G{\"u}rdogan, S. He, T. Lam, J. Trnka for many invaluable discussions about amplituhedra and positive geometries.
The work presented in this review was partially funded by the Deutsche Forschungsgemeinschaft (DFG, German Research Foundation) -- Projektnummern 270039613, 404358295 and 404362017.

%%%%%%%%%%%%%%%%%%%%%%%%%%%%%%%%%%%%%%%%%%%%%%%
%%%%%%%%%%%%%%%%%%%%%%%%%%%%%%%%%%%%%%%%%%%%%%%
%%%%%%%%%%%%%%%%%%%%%%%%%%%%%%%%%%%%%%%%%%%%%%%
%%%%%%%%%%%%%%%%%%%%%%%%%%%%%%%%%%%%%%%%%%%%%%%
%%%%%%%%%%%%%%%%%%%%%%%%%%%%%%%%%%%%%%%%%%%%%%%

\appendix

\section{Kinematic spaces for \texorpdfstring{$\mathcal{N}=4$}{} sYM}
\label{app:spaces}

In this appendix we collect some information on the coordinate spaces which are used throughout the review.

%%%%%%%%%%%%%%%%%%%%%%%%%%%%%%%%%%%%%%%%%%%%%%%

\paragraph{Spinor helicity space and twistor variables.}

In a massless theory  in four dimensions with $p_i^2=0$ for all particles, one can write each momentum as
\begin{equation}
p_i^{a\dot a}=\lambda_i^a\widetilde\lambda_i^{\dot a}\,,
\end{equation} 
in terms of two spinor variables $\lambda$ and $\widetilde\lambda$.
In $\mathcal{N}=4$ SYM, we consider an extension of the spinor helicity space: there are two superspaces on which the theory can be defined
\begin{itemize}
\item chiral superspace $(\lambda^{\alpha},\tilde\lambda^{\dot\alpha}|\eta^A)$: parametrised by Grassmann-odd variables,  $\eta^A$, transforming as a fundamental representation of the $SU(4)$ R-symmetry. This superspace is relevant for the amplituhedron.
\item non-chiral superspace $(\lambda^{\alpha},\eta^a | \tilde\lambda^{\dot\alpha},\tilde\eta^{\dot a})$: parametrised by two sets of Grassmann-odd variables, $\eta^a,\widetilde\eta^{\dot a}$, which both are transforming as fundamental representations of $SU(2)$. One can think of $\widetilde \eta^{\dot a}$ as Fourier conjugate variables to $\eta^{3,4}$. This superspace is relevant for the momentum amplituhedron.
\end{itemize}
From the on-shell chiral superspace, supertwistor variables  are defined as $\mathcal{W}_i^{\mathcal{A}}=(\widetilde \mu^{\alpha}_{i},{\widetilde \lambda}^{\dot\alpha}_{i}| \eta^{A}_{i})$, where $\widetilde \mu^{\alpha}_{i}$ is  the Fourier conjugate to $\lambda_{i}^{\alpha}$. They linearise the action of superconformal symmetry.

%%%%%%%%%%%%%%%%%%%%%%%%%%%%%%%%%%%%%%%%%%%%%%%
 
\paragraph{Dual superspace and momentum twistor variables.}

Starting from the on-shell chiral superspace, one can define another, dual superspace with coordinates $(x,\theta)$ for $i = 1, \ldots , n$ with
\begin{equation}
x^{a\dot a}_i - x^{a\dot a}_{i-1} = \lambda^a_i \widetilde \lambda^{\dot a}_i\qquad \theta^{a A}_i - \theta^{aA}_{i-1} = \lambda^a_i \eta^{ A}_i\,.
\end{equation}
This is the space where the $n$-sided null polygon Wilson loop dual to the $n$-point amplitude is naturally formulated.
The (super) momentum twistors are in the fundamental representation of the superconformal group of this dual space; explicitly
\begin{equation}
\label{eq:momtw}
\mathcal{Z}_i = (z^a_i|\chi^A_i) = (\lambda_{i a}, \mu_i^{\dot a}|\chi^A_i) \equiv (\lambda_{i a}, x^{a\dot a}\lambda_{ia}|\theta_i^{aA}\lambda_{ia})\,.
\end{equation}
\begin{samepage}
The momentum twistors are unconstrained and they determine $\widetilde\lambda,\eta$ via,
\begin{equation}
(\widetilde\lambda|\eta)_i =\frac{\langle i-1\,i\rangle (\mu|\chi)_{i+1}+\langle i+1\,i-1\rangle (\mu|\chi)_{i}+\langle i\,i+1\rangle (\mu|\chi)_{i-1}}{\langle i-1\,i\rangle\langle i\,i+1\rangle}\,.
\end{equation}
 They linearise the action of dual superconformal symmetry.
\end{samepage}

\bibliographystyle{nb}

\bibliography{review_amplituhedra}

\end{document}